\documentclass[sigconf]{acmart}
\usepackage{booktabs}
\usepackage{algorithm}
\usepackage{algorithmicx}
\usepackage{algpseudocode}
\usepackage{multirow}
\usepackage{adjustbox}
\usepackage{amsthm}
\usepackage{enumitem}
\usepackage{colortbl}
\definecolor{gainsboro}{RGB}{ 233,233,233}

\copyrightyear{2023}
\acmYear{2023}
\setcopyright{rightsretained}
\acmConference[SIGIR '23] {Proceedings of the 46th International ACM SIGIR Conference on Research and Development in Information Retrieval}{July 23--27, 2023}{Taipei, Taiwan.}
\acmBooktitle{Proceedings of the 46th International ACM SIGIR Conference on Research and Development in Information Retrieval (SIGIR '23), July 23--27, 2023, Taipei, Taiwan}
\acmPrice{}
\acmISBN{978-1-4503-9408-6/23/07}
\acmDOI{10.1145/3539618.3591725}

\newcommand{\proposed}{\textsf{MELT}}

\begin{document}
\title{MELT: Mutual Enhancement of Long-Tailed User and Item for Sequential Recommendation}

\author{Kibum Kim}
\orcid{0000-0002-7381-019X}

\affiliation{
  \institution{KAIST ISysE}
  \city{Daejeon}
  \state{}
  \country{Republic of Korea}
  \postcode{34141}
}
\email{kb.kim@kaist.ac.kr}

\author{Dongmin Hyun}
\affiliation{%
  \institution{POSTECH PIAI}
  \city{Pohang}
  \state{}
  \country{Republic of Korea}
  \postcode{34141}
  }
\email{dm.hyun@postech.ac.kr}

\author{Sukwon Yun}
\affiliation{%
  \institution{KAIST ISysE}
  \city{Daejeon}
  \state{}
  \country{Republic of Korea}
  \postcode{34141}
}
\email{swyun@kaist.ac.kr}

\author{Chanyoung Park}
\authornote{Corresponding author}
\affiliation{%
  \institution{KAIST ISysE \& AI}
  \city{Daejeon}
  \state{}
  \country{Republic of Korea}
  \postcode{34141}
}
\email{cy.park@kaist.ac.kr}



\begin{abstract}
The long-tailed problem is a long-standing challenge in Sequential Recommender Systems (SRS) in which the problem exists in terms of both users and items.
While many existing studies address the long-tailed problem in SRS, they only focus on either the user or item perspective.
However, we discover that the long-tailed user and item problems exist at the same time, and considering only either one of them leads to sub-optimal performance of the other one.
In this paper, we propose a novel framework for SRS, called 
\textbf{M}utual \textbf{E}nhancement of \textbf{L}ong-\textbf{T}ailed user and item (\proposed{}), that jointly alleviates the
long-tailed problem in the perspectives of both users and
items.~\proposed~consists of bilateral branches each of which is responsible for long-tailed users and items, respectively, and the branches are trained to mutually enhance each other, which is trained effectively by a curriculum learning-based training. \proposed{} is model-agnostic in that it can be seamlessly integrated with existing SRS models. Extensive experiments on eight datasets demonstrate the benefit of alleviating the long-tailed problems in terms of both users and items even without sacrificing the performance of head users and items, which has not been achieved by existing methods. 
To the best of our knowledge,~\proposed~is the first work that jointly alleviates the long-tailed user and item problems in SRS.
Our code is available at {\href{https://github.com/rlqja1107/MELT}{https://github.com/rlqja1107/MELT}}.
\vspace{-1ex}
\end{abstract}

\maketitle

\section{Introduction}
\begin{figure}[t]
\vspace{3ex}
    \begin{center}
        \includegraphics[width=1\columnwidth]{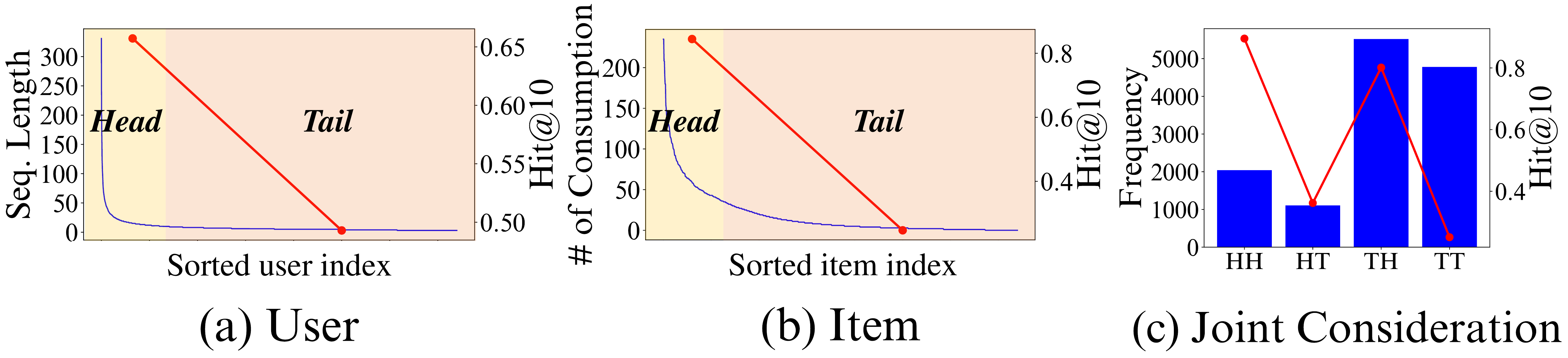}
    \end{center}
    \vspace{-2ex}
  \caption{Hit@10 performance (red) according to (a) the lengths of users' consumption sequences, (b) the number of consumed items, and (c) the combination of head/tail users and head/tail items on Amazon Music data.}
    \label{fig:fig_1}
    \vspace{-3ex}
\end{figure}

Despite the success of recommender systems, they have suffered from chronic long-tailed problems in terms of users and items.
The long-tailed user problem, which is prevalent in online services, refers to the situation in which users with few interactions (i.e., tail users) greatly outnumber users with many interactions (i.e., head users) \cite{TP,ASReP}. 
We observe in Figure~\ref{fig:fig_1}(a) that in a real-world dataset (i.e., Amazon Music data), the number of tail users greatly outnumbers that of head users, while the performance on tail users is significantly lower than that on head users. 
Hence, it is crucial to focus on improving the performance on tail users as the number of active users therein can be eventually increased, which leads to an increase in revenue. 
On the other hand, the long-tailed item problem refers to the situation in which users mainly consume few popular items (i.e., head items) compared with many unpopular items (i.e., tail items), which leads to recommender models being biased towards head items although most items are tail items (Figure~\ref{fig:fig_1}(b)).
In this regard, recommending appropriate tail items is vital as online services can make users stay on their systems by increasing the serendipity of their recommendation, which eventually increases the revenue \cite{TailNet,yin2012challenging}.

Recent studies focus on sequential recommender systems (SRS) to address the long-tailed problem \cite{ASReP,DTSR,TP,CITIES,TailNet,INSERT,yun2022lte4g} as considering the sequential information of interacted items is helpful for alleviating the long-tail problem \cite{kim2019sequential}.
Specifically, recent works address the long-tailed user problem by augmenting users' historical interaction \cite{arxiv,ASReP} or performing adversarial training to map the head and tail users into a shared latent space \cite{TP}. However, they only focus on improving the performance of tail users, while ignoring the long-tailed item problem, which results in a poor performance on tail items.
On the other hand, other works address the long-tailed item problem by enhancing the tail item representation \cite{CITIES} or adjusting the final prediction with attention scores derived by the two types (i.e, head and tail) of items \cite{TailNet}. However, they overlook the long-tailed user problem resulting in a poor performance on tail users.

In this work, we argue that \textit{jointly addressing both long-tailed user and item problems is essential in practice}. 
To corroborate our argument, we split the users into four groups considering the long-tailedness of both users and items, and show the frequency along with the recommendation performance in terms of Hit@10 for each of the subsets in Figure~\ref{fig:fig_1}(c)\footnote{SASRec~\cite{sasrec} is used for the experiments.}.
For example, the HT group indicates Head users whose last consumption is a Tail item\footnote{We follow leave-one-out protocol, i.e., we aim to predict the last item each user is likely to consume. We follow the Pareto principle \cite{pareto,box1986analysis} for splitting users and items.}
As expected, the recommendation performance on the TT group is the lowest among all the groups. However, we observe that the number of users in the TT group accounts for a significant portion (i.e., 38\%), which implies that the performance on TT is crucial and should not be overlooked.

A straightforward approach for jointly addressing the long-tailed user and item problems would be to naively combine the two types of existing approaches, i.e., one that only tackles the long-tailed user problem (e.g., ASReP \cite{ASReP}), and the other one that only tackles the long-tailed item problem (e.g., CITIES \cite{CITIES}). In Table~\ref{tab:intro}, we compare the performance of ASReP, CITIES, and a naively combined version of CITIES and ASReP on the TT group\footnote{The performance of HH, HT, and TH groups are shown in Section~\ref{sec:fine_grained}.}.
We find that although the naive combination outperforms ASReP, it rather performs worse than CITIES, which is unexpected.
The main reason is two-fold: 1) items augmented by ASReP are mostly head items thereby aggravating the long-tailed item problem, and 2) the augmentation model of ASReP, which is fixed after being trained, cannot reflect the representations of tail items that are iteratively updated by CITIES.
That is, the naive combination of ASReP and CITIES fails to mutually enhance each other as they are independently considered.
Furthermore, naively combining two separate models  linearly increases the model complexity in terms of both time and space, which is impractical to be applied to real-world applications.

\begin{table}[t]
\caption{Performance on TT group in terms of Hit@10. For \textit{ASReP+CITIES}, we train CITIES on 
users’ consumption sequences each of which is augmented by ASReP.}
\resizebox{\columnwidth}{!}{
\begin{tabular}{l|cccc}
\toprule
\multirow{2}{*}{\textbf{Model}} & \multicolumn{4}{c}{\textbf{Hit@10 on TT}}                                                                 \\ \cline{2-5} 
                                 & \textbf{Music}                             & \textbf{Beauty}                           & \textbf{Automotive}                        & \textbf{Behance}                           \\ \midrule
ASReP                                                                                                                                     & 0.246                                      & 0.160                                     & 0.114                                      & 0.287                                      \\
CITIES                                                                                                                                    & 0.279                                      & 0.191                                     & 0.130                                      & 0.298                                      \\\midrule\midrule
ASReP+CITIES                                                                                                                       & 0.273 (\textcolor{red}{-2.2\%})            & 0.169 (\textcolor{red}{-11.5\%})          & 0.124 (\textcolor{red}{-4.6\%})            & 0.296 (\textcolor{red}{-0.7\%})            \\ \midrule
\proposed                                                                                                                           & \textbf{0.312 (\textcolor{blue}{+11.8\%})} & \textbf{0.197 (\textcolor{blue}{+3.1\%})} & \textbf{0.149 (\textcolor{blue}{+14.6\%})} & \textbf{0.371 (\textcolor{blue}{+24.5\%})} \\ \bottomrule
\end{tabular}

}
\label{tab:intro}
\vspace{-2ex}
\end{table}

In this paper, we propose a simple but effective SRS, called \textbf{M}utual \textbf{E}nhancement of \textbf{L}ong-\textbf{T}ailed user and item (\proposed), that jointly alleviates the long-tailed user and item problems. Our proposed framework consists of bilateral branches each of which is responsible for long-tailed users and items, respectively. 
The main idea is to train two embedding generators in each branch based on head users and head items, and iteratively enhance the representation of tail users and tail items in an end-to-end manner.
More precisely, in the user branch, an embedding generator is trained to generate a head user's complete representation given the incomplete representation obtained from the user's sampled subsequence as the input. 
On the other hand, in the item branch, another embedding generator is trained to generate a head item's complete representation given the incomplete representation obtained from the item's partial interactions, i.e., sampled sequences each of which ends with that item.
Having trained the two embedding generators, we update the representation of tail users and items, and then train the generators again based on the updated representations, which eventually enhances the user/item representations in general.
For an effective training, we devise a curriculum learning (CL) strategy that trains the model from easy (i.e., a large number of interactions regarding users and items) to hard (i.e., a small number of interactions regarding users and items), which stabilizes the knowledge-transfer process from head users and head items.

Extensive experiments on \textit{eight} real-world benchmark datasets show that~\proposed~significantly improves the performance on both the tail users and tail items compared with baseline methods. A further appeal of \proposed{} is that it does not sacrifice the performance on head users and head items,
which has not been achieved by existing methods. It is important to note that \proposed{} is a model-agnostic framework that is applicable to any existing SRSs to jointly address the long-tailed problem in terms of both users and items.

Our contributions are summarized as follows:
\begin{itemize}[leftmargin=5mm]
    \item We propose a simple but effective metod, called~\proposed, that is designed to jointly alleviate the long-tailed user and item problems in sequential recommendation.
    
    \item We design bilateral branches to address the long-tailed user and item problems, while making the branches mutually enhance each other.
    \item \proposed{} is a model-agnostic framework that is applicable to any
SRS to address the long-tailed problems.
    \item Extensive experiments show that~\proposed~enhances the performance of tail users and tail items without sacrificing the performance of both head users and head items.
\end{itemize}


\section{Preliminary}
Let $\mathcal{U}$ and $\mathcal{I}$ denote the set of users and items, respectively. A user $u\in\mathcal{U}$ has a sequence of item consumption sorted by the timestamp denoted by $\mathcal{S}_u=\left [ i_1^u, i_2^u,..., i^u_{|\mathcal{S}_{u}|} \right ]$, where $|\mathcal{S}_u|$ is the sequence length and $i_{t}^u\in\mathcal{I}$ is the $t$-th item in the sequence. 
We denote the representations of user $u$ and item $i$ by $p_u \in \mathbb{R}^d$ and $q_i \in \mathbb{R}^d$, respectively, where $d$ is the hidden dimension size.
The goal of the SRS task is to predict the user's next consumption $i^u_{|\mathcal{S}_{u}|+1}$ based on the user's given sequence $\mathcal{S}_u$, and a sequence encoder $f_\theta(\cdot)$, which produces the representation of an item sequence $S_u$, i.e., $r_u=f_{\theta}(\mathcal{S}_u)\in\mathbb{R}^d$.
It is important to note that for SRS models that obtain a user's representation by encoding the user's item sequence, e.g., SASRec~\cite{sasrec}, rather than explicitly training an independent user embedding vector, the representation for a user $u$ is obtained by $p_u=r_u=f_{\theta}(\mathcal{S}_u)$.

Note that any other sequence encoder such as BERT4Rec \cite{bert4rec}, FMLP \cite{FMLP}, and GRU4Rec \cite{gru4rec} can be used instead of SASRec. Although we mainly use SASRec as the sequence encoder throughout the paper, we later show the result of using FMLP to verify that our proposed framework is model-agnostic (Table~\ref{tab:fmlp}). We summarize the notations used throughout the paper in Table~\ref{tab:notation}.

\smallskip
\noindent\textbf{Splitting Users and Items into Head and Tail. }
\label{sec:ht_split}
Following existing studies \cite{TP,TailNet,CITIES}, we sort the users according to their sequence lengths in descending order, and consider top $\alpha\%$ users as head users (i.e., $u^H\in\mathcal{U}^H$) and the remaining users as tail users (i.e., $u^T\in\mathcal{U}^T$), where $\mathcal{U}^H$ and $\mathcal{U}^T$ are sets of head and tail users, respectively. Similarly, we set the top $\alpha\%$ frequently-interacted items as head items (i.e., $i^H\in\mathcal{I}^H$) and the remaining items as tail items (i.e., $i^T\in\mathcal{I}^T$), where $\mathcal{I}^H$ and $\mathcal{I}^T$ are sets of head and tail item, respectively. 
Note that the well-known Pareto Principle \cite{pareto,box1986analysis} is when $\alpha=20\%$.

\section{Method}
\begin{figure}[t]
    \begin{center}
        \includegraphics[width=0.65\columnwidth]{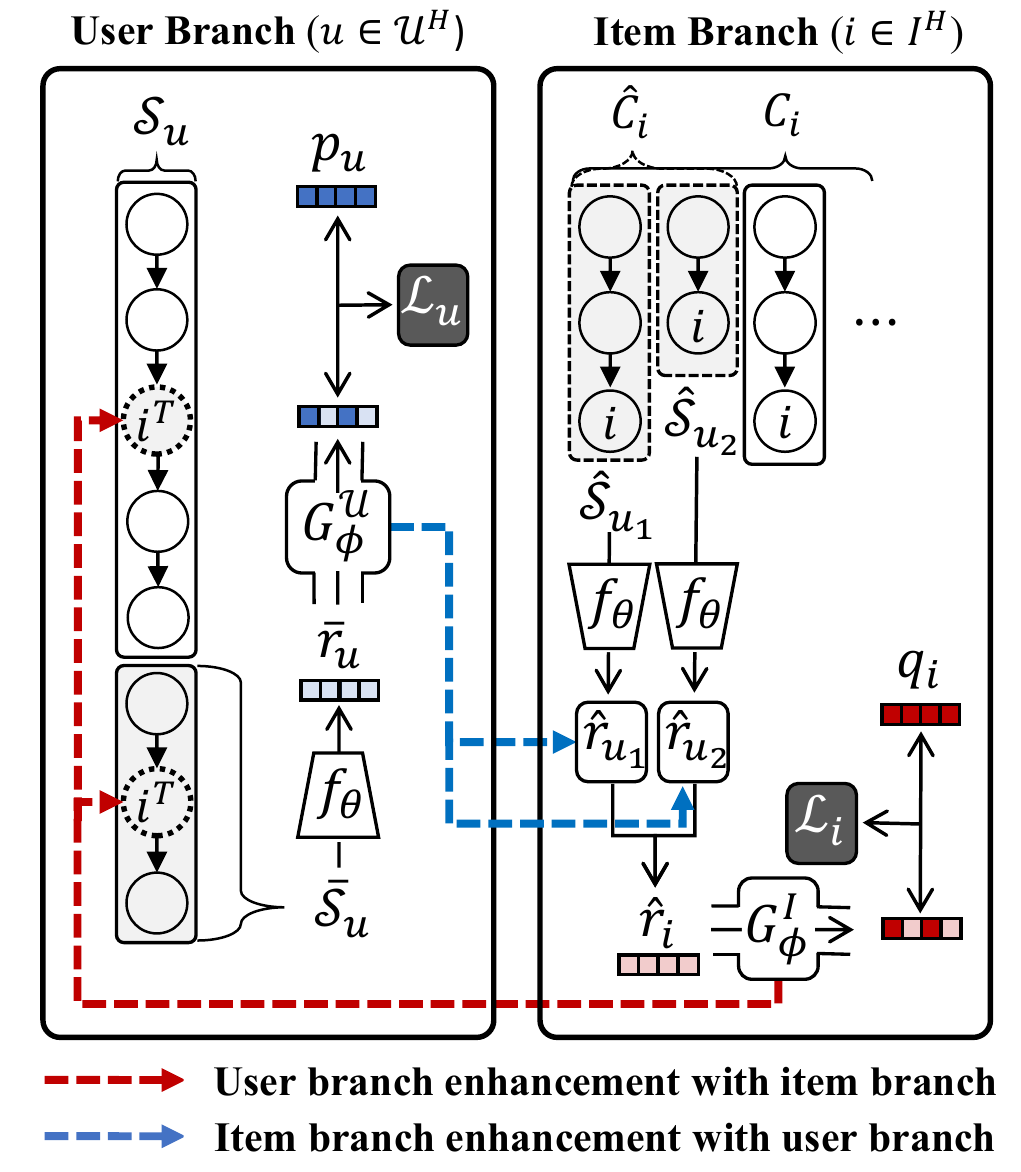} 
    \end{center}
    \vspace{-2ex}
    \caption{Overall framework of user and item branch.}
    \label{fig:model}
    \vspace{-3ex}
\end{figure}

This section describes bilateral branches for long-tailed users (\textbf{Section \ref{sec:bbu}}) and items (\textbf{Section~\ref{sec:bbi}}), and how the two branches are coupled in a mutually enhancing manner (\textbf{Section~\ref{sec:jt}}).
Then, we introduce the curriculum learning strategy to stabilize the knowledge transfer process (\textbf{Section~\ref{sec:cl}}) followed by the description of the model training/inference process (\textbf{Section~\ref{sec:model_train}}). 
For simplicity, we denote a head user and a head item by $u$ and $i$, instead of $u^H$ and $i^H$. For tail cases, we explicitly denote tail users and tail items as $u^T$ and $i^T$, respectively. Figure~\ref{fig:model} shows the overall architecture of~\proposed.

\begin{table}[t]
\caption{Notations}
\vspace{-2ex}
\resizebox{0.95\linewidth}{!}{
\begin{tabular}{l|ll}
\toprule
\textbf{Notation}           & \multicolumn{2}{l}{\textbf{Description}}                                         \\ \midrule \midrule
$\mathcal{U}, \mathcal{I}$             & \multicolumn{2}{l}{User set, item set}                                         \\
$\mathcal{U}^H$,$\mathcal{U}^T$,$\mathcal{I}^H$,$\mathcal{I}^T$             & \multicolumn{2}{l}{Set of Head user, Tail user, Head item, Tail item}                                         \\
$\mathcal{S}_u$             & \multicolumn{2}{l}{Sequence of user $u$}                                         \\
$f_{\theta}$                & \multicolumn{2}{l}{Sequence encoder (e.g., SASRec)}                              \\
\midrule
$\bar{\mathcal{S}}_u$       & \multicolumn{2}{l}{Truncated subsequence of user $u$ that contains recent interactions}                               \\
$\bar{r}_{u}$             & \multicolumn{2}{l}{User $u$'s subsequence representation ($\bar{r}_{u}=f_{\theta}(\bar{\mathcal{S}}_u)$)}   \\
$p_u$                       & \multicolumn{2}{l}{User $u$'s complete representation ($p_u=r_u=f_{\theta}(\mathcal{S}_u)$ for SASRec)}                           \\
${p}_u^+$                       & \multicolumn{2}{l}{Enhanced user $u$'s representation}                           \\
$G_\phi^{\mathcal{U}}$ & \multicolumn{2}{l}{User embedding generator} \\
\midrule
$\mathcal{C}_i$             & \multicolumn{2}{l}{Set of users' subsequences that end with that item $i$}          \\
$\hat{\mathcal{C}}_i$               & \multicolumn{2}{l}{Randomly sampled users' subsequences from $\mathcal{C}_i$}       \\
$\hat{r}_{u}$             & \multicolumn{2}{l}{User $u$'s subsequence representation that ends with an item}   \\
$\hat{r}_{i}$             & \multicolumn{2}{l}{Contextualized representation of item $i$ obtained from $\hat{\mathcal{C}}_i$ set}   \\
$q_i$                       & \multicolumn{2}{l}{Item $i$'s complete representation}                           \\ 

${q}_i^+$                       & \multicolumn{2}{l}{Enhanced item $i$'s representation}                           \\
$G_\phi^{\mathcal{I}}$ & \multicolumn{2}{l}{Item embedding generator}  \\ 
\bottomrule

\end{tabular}
}
\label{tab:notation}
\vspace{-1.5ex}
\end{table}

\subsection{User Branch}
\label{sec:bbu}
To alleviate the lack of interactions of tail users,
we leverage the information gap between the head user's representation (i.e., $p_u\in \mathbb{R}^d$) obtained based on the user's entire item consumption history, and the user's representation obtained from the user's partial interactions. Concretely, we deliberately truncate a head user's entire item sequence  $\mathcal{S}_{u}$ into a subsequence $\bar{\mathcal{S}}_{u}$ that contains $R$ recent interactions: 
\begin{equation}
\small
\bar{\mathcal{S}}_u=\left [i_{|\mathcal{S}_u|-R+1}^u, i_{|\mathcal{S}_u|-R+2}^u,...,i_{|\mathcal{S}_u|}^u  \right].
\end{equation}
Note that $\bar{\mathcal{S}}_u$ plays a role as the item consumption sequence of a tail user aiming at simulating the situation in which a user lacks interactions.
Next, the model learns to generate the head user's representation $p_u$ given $\bar{S}_{u}$, and then transfer the learned knowledge to existing tail users (i.e., $u^T$). 
Formally, we adopt a sequence encoder $f_\theta{(\cdot)}$ to produce a head user's subsequence representation  (i.e., $\bar{r}_u$):
\begin{equation}
\small
\bar{r}_{u}=f_{\theta}(\bar{\mathcal{S}}_{u})
\end{equation}
where $\bar{r}_{u}$ captures the user's recent interest.
Given $\bar{r}_u \in \mathbb{R}^d$, we aim to generate a user representation that is close to the head user's complete representation (i.e., $p_u$) by minimizing the following loss:
\begin{equation}
\small
\label{eqn:eq1}
\mathcal{L}_{u}=\left\| p_u-G_\phi^{\mathcal{U}}(\bar{r}_u)\right\|^2  
\end{equation}
where $G_\phi^{\mathcal{U}}:\mathbb{R}^d \rightarrow \mathbb{R}^d$ is a user embedding generator, which is responsible for generating the complete representation based on the input $\bar{r}_u$. 
By minimizing Equation~\ref{eqn:eq1}, we expect $G_\phi^{\mathcal{U}}$ to contain sufficient knowledge to generate a user's complete representation given a short item consumption sequence, which can be used to enhance the representation of a tail user $p_{u^T}$ as follows:
\begin{equation}
\small
\label{eqn:usertransfer}
{p}_{u^T}^+ =G_\phi^{\mathcal{U}}(r_{u^T}) + \beta \, p_{u^T}
\end{equation}
where ${p}_{u^T}^+$ is the enhanced representation of the tail user $p_{u^T}$, $r_{u^T}=f_{\theta}(\mathcal{S}_{u^T})\in\mathbb{R}^d$ is a tail user's representation obtained based on the user's entire item consumption sequence $\mathcal{S}_{u^T}$, and $\beta\in[0,1]$ is a hyperparameter that controls the contribution of the tail user's original representation $p_{u^T}$.
By considering the representation generated based on the knowledge obtained from head users through $G_\phi^{\mathcal{U}}$ (i.e., $G_\phi^{\mathcal{U}}(r_{u^T})\in\mathbb{R}^d$) in addition to the original representation $p_{u^T}$, we argue that the low-quality representation of tail users incurred by their lack of item consumption history can be supplemented.

\subsection{Item Branch}
\label{sec:bbi} 
To alleviate the lack of interactions of tail items, we leverage the information gap between the head item's representation (i.e., $q_i\in \mathbb{R}^d$) obtained based on the item's entire interactions with users, and the item's representation obtained from the item's partial interactions. We first define a set of interactions regarding item $i$, i.e., $\mathcal{C}_i$, as a set of users' subsequences that end with the item $i$ such that:
\begin{equation}
\small
\mathcal{C}_i=\{\hat{\mathcal{S}}_u|\, \hat{\mathcal{S}}_{u, |\hat{\mathcal{S}}_u|} = i, \, \hat{\mathcal{S}}_u \prec   \mathcal{S}_u, \forall u\in\mathcal{U} \}    
\label{eqn:ci}
\end{equation}
where $\hat{\mathcal{S}}_u$ is a user's item consumption subsequence that is truncated from the first consumed item up to item $i$ in the original sequence $\mathcal{S}_u$, $\hat{\mathcal{S}}_{u, |\hat{\mathcal{S}}_u|}$ indicates the last item in the truncated sequence, and $\prec$ indicates the subsequence relation.
For example, given two item sequences $[i_1, i_2, i_4]$ and $[i_1, i_3, i_2, i_4]$, the set of subsequences regarding item $i_2$ is defined as $\mathcal{C}_2=\{[i_1, i_2], [i_1, i_3, i_2]\}$.
By passing a subsequence that ends with item $i$ through the sequence encoder (i.e., $f_\theta(\cdot)$), we obtain a contextualized representation of item $i$.

We simulate the information gap by randomly sampling $K$ subsequences ($\hat{\mathcal{C}}_{i}$) from the set of interactions ($\mathcal{C}_{i}$), i.e., $|\hat{\mathcal{C}}_{i}| = K, \, \hat{\mathcal{C}}_{i} \subset \mathcal{C}_{i}$.
Then, the model learns to generate the head item's complete representation (i.e., $q_i$) from the sampled set of interactions (i.e., $\hat{\mathcal{C}}_i$), and transfer the learned knowledge to existing tail items.
Formally, we represent item $i$ based on $\hat{\mathcal{C}}_i$ as follows:
\begin{equation}
\small
\hat{r}_{i}=\frac{1}{K}\sum_{\hat{\mathcal{S}}_u \in \hat{\mathcal{C}}_{i}} \hat{r}_u, \quad \hat{r}_u = f_\theta(\hat{\mathcal{S}}_u)  
\label{eqn:avg}
\end{equation}
Given the representation of item $i$ (i.e., $\hat{r}_{i}$) computed based on its incomplete set of interactions (i.e., $\hat{\mathcal{C}}_i$), we aim to generate its complete representation (i.e., $q_i$) by minimizing the following loss:
\begin{equation}
\small
\label{eqn:eq3}
\mathcal{L}_{i}=\left\| q_i-G_\phi^{\mathcal{I}}(\hat{r}_{i})\right\|^2 
\end{equation}
where $G_\phi^{\mathcal{I}}:\mathbb{R}^d \rightarrow \mathbb{R}^d$ is an item embedding generator, which is responsible for generating the complete representation based on the input $\hat{r}_{i}$. 
By minimizing Equation~\ref{eqn:eq3}, we expect $G_\phi^{\mathcal{I}}$ to contain sufficient knowledge to generate an item's complete representation, which can be used to enhance the representation of a tail item $q_{i^T}$ as follows:
\begin{equation}
\small
\label{eqn:itemtransfer}
{q}_{i^T}^+=G_\phi^{\mathcal{I}}(r_{i^T}) + \gamma  \, q_{i^T}
\end{equation}
where ${q}_{i^T}^+$ is the enhanced representation of the tail item $q_{i^T}$, ${r}_{i^T}=\frac{1}{|\mathcal{C}_{i^T}|}\sum_{\hat{\mathcal{S}}_u \in {\mathcal{C}}_{i^T}} f_\theta(\hat{\mathcal{S}}_u)\in\mathbb{R}^d$ is the tail item's representation obtained based on its entire set of interactions (i.e., $\mathcal{C}_{i^T}$), and
$\gamma \in [0,1]$ is a hyperparameter that controls the contribution of the tail item’s original representation $q_{i^T}$.
By considering the representation generated based on the knowledge obtained from head items through $G_\phi^{\mathcal{I}}$ (i.e., $G_\phi^{\mathcal{I}}(r_{i^T})\in\mathbb{R}^d$) in addition to the original representation $q_{i^T}$, we argue that the low-quality representation of tail items incurred by their lack of interactions with users can be supplemented.

\subsection{Mutual Enhancement of Bilateral Branches}
\label{sec:jt}
Now that we have independently obtained the user and item branches as described so far, we connect the branches so as to facilitate the mutual enhancement between them, which jointly alleviates the long-tailed user and item problems.
First, the \textbf{user branch} utilizes the knowledge contained in the item embedding generator ${G}_\phi^{\mathcal{I}}$ obtained from the item branch (red arrow from ${G}_\phi^{\mathcal{I}}$ in Figure \ref{fig:model}). 
Note that in the user branch, before feeding the sequence of user $u$ (i.e., $\mathcal{S}_u=[i_1, i_2, \cdots, i_{|\mathcal{S}_u|}]$) to the sequence encoder $f_\theta(\cdot)$, each item in the sequence is converted to a $d$-dimensional representation as follows:
\begin{equation}
\small
    E_u = [q_{i_1}, q_{i_2}, \cdots, q_{i_{|\mathcal{S}_u|}}]
\end{equation}
where $E_u \in \mathbb{R}^{|\mathcal{S}_u| \times d}$ is the representation matrix of items that appear in the user's consumption sequence $\mathcal{S}_u$. We update the representations of tail items in the user's sequence based on the knowledge obtained from the item branch, i.e., ${G}_\phi^{\mathcal{I}}$, as follows:
\begin{equation}
\small
    q^+_i=\begin{cases}
    {G}_\phi^{\mathcal{I}}(r_{i}) + \gamma  \, q_{i} & i \in \mathcal{I}^T,\\
    q_i & \text{otherwise},
  \end{cases}
  \label{eq:eq10}
\end{equation}
where $q_i^+ \in \mathbb{R}^d$ is the enhanced representation of item $i$, and $r_i = \frac{1}{|\mathcal{C}_{i}|}\sum_{\hat{\mathcal{S}}_u \in \mathcal{C}_{i}} \hat{r}_u$ is the contextualized representation of item $i$. 
Note that we only update the representation of tail items (dashed circles in Figure \ref{fig:model}), while the representation of head items  (solid circles in Figure \ref{fig:model}) is not updated.
We hence obtain an enhanced representation matrix as follows:
\begin{equation}
\small
    E_u^+ = [q_{i_1}^+, q_{i_2}^+, \cdots, q^+_{i_{|\mathcal{S}_u|}}].
\end{equation}
We argue that the above procedure enhances the overall quality of item representations contained in user $u$'s consumption sequence, which in turn enhances the quality of the user representation that is computed based on user $u$'s sequence, since the representation of items in the sequence is enhanced.
Moreover, it is important to note that the enhanced $E^{+}_{u}$ eventually 
helps improve the knowledge of ${G}_\phi^{\mathcal{U}}$ in 
Equation~\ref{eqn:eq1}, thereby alleviating the long-tailed user problem.

Similarly, the \textbf{item branch} utilizes the knowledge contained in the user embedding generator ${G}_\phi^{\mathcal{U}}$ obtained from the user branch (blue arrow from ${G}_\phi^{\mathcal{U}}$ in Figure \ref{fig:model}). Recall that in the item branch, item $i$ is represented by the subsequences of users who consumed item $i$ (i.e., $\hat{\mathcal{S}}_u \in \hat{\mathcal{C}}_i$). We thus update user $u$'s representation $\hat{r}_u$ obtained from the subsequence $\hat{\mathcal{S}}_u$ based on the knowledge obtained from the user branch, i.e., ${G}_\phi^{\mathcal{U}}$, as follows: 
\begin{equation}
\small
\label{eqn:eq12}
    \hat{r}^+_u={G}_\phi^{\mathcal{U}}(\hat{r}_u) + \beta \, p_u 
\end{equation}
where $\hat{r}_u^+ \in \mathbb{R}^d$ is the enhanced representation of user $u$
(dashed sequences in the item branch in Figure \ref{fig:model}), and $\hat{r}_u = f_\theta(\hat{\mathcal{S}}_u)$. Recall that in the item branch, each user’s sequence is truncated into a subsequence regardless of user belonging to
head or tail as described in Section~\ref{sec:bbi}. Hence, we update the
representations of all subsequences based on Equation~\ref{eqn:eq12}. Since the representation of item $i$ (i.e., $\hat{r}_{i}$) is obtained by an average of $\hat{r}^+_u$ given $\hat{\mathcal{S}}_u \in \hat{\mathcal{C}}_i$ as shown in Equation~\ref{eqn:avg}, we argue that the above procedure enhances the overall quality of user representations, which in turn enhances the quality of representations computed based on the user representations.
Note that we only update the representation of tail items (i.e., $\bar{r}_{i^T}$), as the increase in the model complexity
outweighs the increase in the recommendation performance.
Moreover, it is important to note that the enhanced $\hat{r}_{i}$ eventually 
helps improve the knowledge of ${G}_\phi^{\mathcal{I}}$ in
Equation~\ref{eqn:eq3}, thereby alleviating the long-tailed item problem.

It is worthwhile to mention that the mutual enhancement process between the bilateral branches described so far occurs iteratively in an end-to-end manner. Specifically, the enhanced knowledge of ${G}_\phi^{\mathcal{U}}$ from the item branch further helps to improve the knowledge contained in ${G}_\phi^{\mathcal{I}}$, while the more enhanced ${G}_\phi^{\mathcal{I}}$'s knowledge further helps to improve the ${G}_\phi^{\mathcal{U}}$'s knowledge. The iterative mutual enhancement process gradually encourages the knowledges be optimal ones, resulting in further alleviating the long-tailed user and item problems.

\subsection{Curriculum Learning}
\label{sec:cl}
Since the knowledge is transferred from head users/items to tail users/items in our proposed framework, the model performance depends on the quality of head users'/items' representation, which further depends on the number of associated interactions therein.
Specifically, we propose a curriculum learning (CL) strategy that gradually trains the model from easy (i.e., large $|\mathcal{S}_u|$, $|\mathcal{C}_i|$) to hard (i.e., small $|\mathcal{S}_u|$, $|\mathcal{C}_i|$). 
The CL \cite{bengio2009curriculum} strategy mimics the human's learning process and can eventually enhance the knowledge obtained from head users and head items by stabilizing the knowledge-transfer process.
We mainly describe the CL strategy for the user branch, but the approach is identically applicable to the item branch.

For the \textbf{user branch}, inspired by the cosine annealing \cite{loshchilov2016sgdr}, we control the weight of each user in the loss based on the difficulty of learning knowledge from head users:
\vspace{-1ex}
\begin{equation}
    \mathcal{L}_{u}= w_u \left\| p_u-G^{\mathcal{U}}_{\phi}( \bar{r}_u)\right\|^2 
\end{equation}
\begin{equation}
\label{eqn:eq5}
w_u=\mathrm{sin}(\frac{\pi}{2}\cdot\frac{e}{e_{max}}+\frac{\pi}{2}\cdot\frac{|S_u|-L_{min}}{L_{max}-L_{min}})
\end{equation}
where $w_u$ is the loss coefficient for a head user $u$, $e$ is the current epoch, $e_{max}$ is the maximum epoch, $L_{min}$ and $L_{max}$ are the minimum and maximum $|S_u|$ in the training data, respectively. At the early stage of training (i.e., $e \ll e_{max}$), 
the model learns more from head users with longer sequence lengths than from head users with shorter sequence lengths. Then, as the training proceeds, the model starts to learn more from head users with shorter sequence lengths than from the head users with longer sequence lengths.

For the \textbf{item branch}, we measure the difficulty based on the number of users' subsequences associated with an head item $i$, i.e., $|\mathcal{C}_i|$, and train the model from easy case (i.e., large $|\mathcal{C}_i|$) to hard case (i.e., small $|\mathcal{C}_i|$). Based on the above Equation \ref{eqn:eq5}, we obtain the loss coefficient $w_i$ for each head item $i$ by replacing $|S_u|$ with $|\mathcal{C}_i|$, and setting $L_{min}$ and $L_{max}$ to the minimum and maximum number of $|\mathcal{C}_i|$ in the training data, respectively.

\begin{figure}[t]
\vspace{-1ex}
    \begin{center}
        \includegraphics[width=0.95\columnwidth]{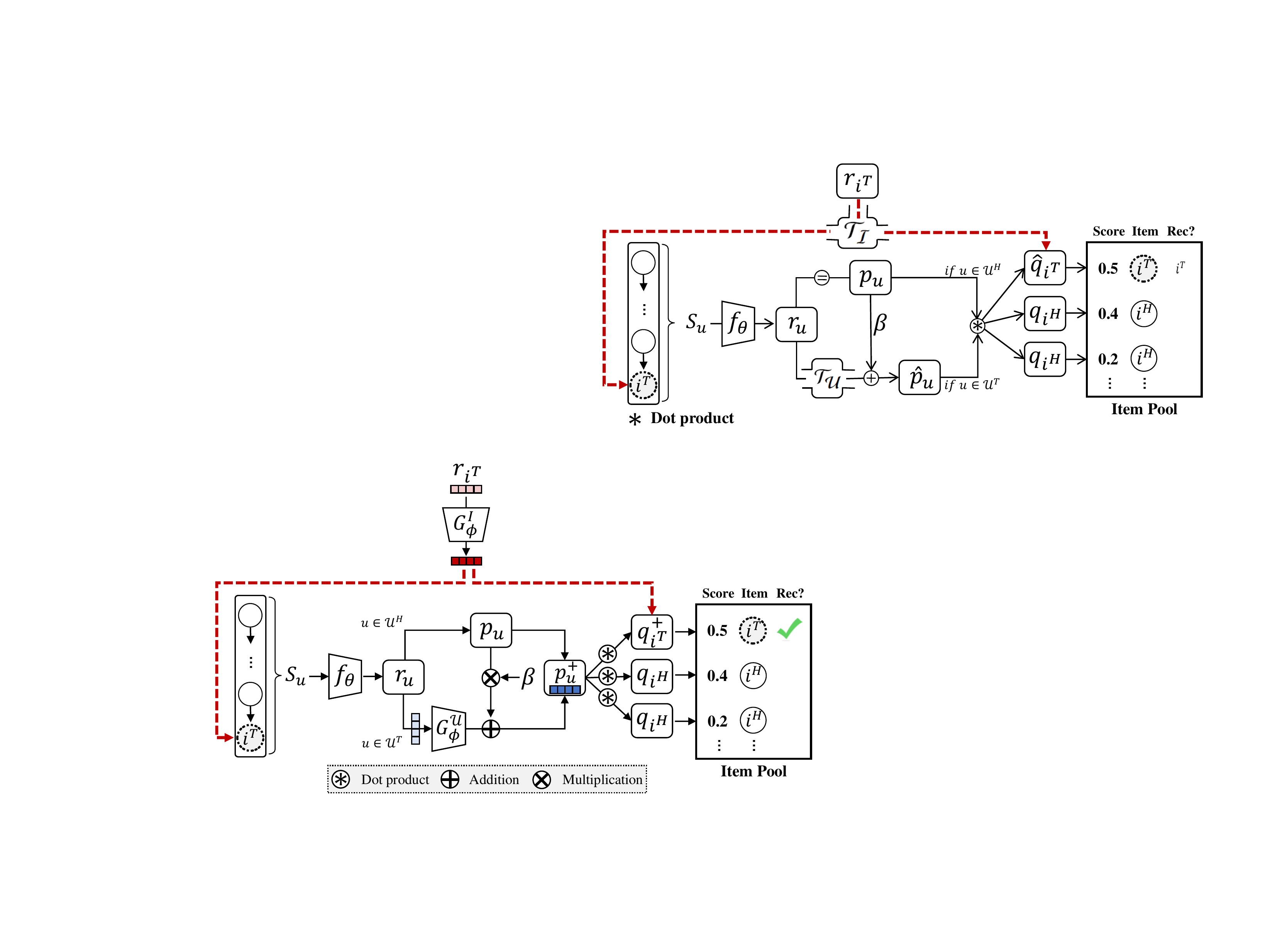}
    \end{center}
    \vspace{-2ex}
    \caption{Overview of inference phase.}
    
    \label{fig:infer}
    \vspace{-3ex}
\end{figure}

\subsection{Model Training and Inference}
\label{sec:model_train}
\proposed~is \textbf{train}ed based on the following loss containing parameters for the embedding generators (i.e., $G_\phi^\mathcal{U}$ and $G_\phi^\mathcal{I}$) and the sequence encoder (i.e., $f_\theta$) as follows:
\begin{equation}
\small
\mathcal{L}_{\text{final}}=\lambda_\mathcal{U} \sum_{u\in \mathcal{U}^H}  \mathcal{L}_{u}+\lambda_\mathcal{I}\sum_{i\in \mathcal{I}^H}  \mathcal{L}_{i} +  \mathcal{L}_\text{rec}
\label{eqn:eq6}
\end{equation}
where $\lambda_\mathcal{U}$ and $\lambda_\mathcal{I}$ are hyperparameters for the losses ($\mathcal{L}_u \text{ and } \mathcal{L}_i$), and $\mathcal{L}_\text{rec}$ is the loss of the SRS models, e.g., next item prediction. Note that~\proposed~is a model-agnostic framework that is applicable to any SRS models to address the long-tailed problems.

\looseness=-1
At the \textbf{inference} phase, we first enhance the representation of tail items $i^T$ in the user sequence $\mathcal{S}_u$ (left part of Figure~\ref{fig:infer}). Then, for a head user $u^H$, we obtain the user representation $p_{u^H}=r_{u^H}=f_\theta(\mathcal{S}_{u^H})$, while for a tail user $u^T$, we obtain the enhanced user representation ${p}_{u^T}^+={G}_\phi^{\mathcal{U}}(r_{u^T})+\beta p_{u^T}$. Assuming that we have a candidate pool of items to be recommended for each user, we compute the dot product between the representation of each item in the pool and the user representation obtained above (i.e., ${p}_{u^H}$ for a head user, and ${p}_{u^T}^+$ for a tail user). Note that for a head item $i^H$, we use the learned item representation $q_{i^H}$, while for a tail item $i^T$, we obtain the enhanced item representation ${q}_{i^T}^+={G}_\phi^{\mathcal{I}}(r_{i^T})+\gamma q_{i^T}$.
Finally, we recommend the top-K items based on the dot product scores.

\smallskip
\noindent\textbf{Complexity Analysis of~\proposed. }
We compare the complexity of \proposed{} with the naively combined model (i.e., ASReP+CITIES shown in Table~\ref{tab:intro}) to further demonstrate the efficiency of \proposed{}. For space complexity, \proposed{} needs only $O(2d^2+|\theta|)$ since we use two single feed-forward neural networks for the embedding generators (i.e., $G_\phi^\mathcal{U}$ and $G_\phi^\mathcal{I}$) in addition to the sequence encoder (i.e., $f_\theta$). For time complexity, the bottleneck of the training time of \proposed{} is the extraction of users' subsequences in the item branch (Equation \ref{eqn:ci}), which needs $O(|\mathcal{U}|\times \underset{u\in\mathcal{U}}{max}(|\mathcal{S}_{u}|))$. However, as the users' subsequences are stored in the form of \textit{key-value} dictionary, where \textit{key} and \textit{value} represent the item and the set of subsequences associated the item, respectively. Therefore, it is not necessary to repeatedly extract the users' subsequence during the training stage, which accelerates the training time.

Compared to $O(2d^2+|\theta|)$ space complexity of \proposed{}, the space complexity of naively combined model is $O(13d^2+2|\theta|)$, which consists of sequence encoders for each ASReP and CITIES (i.e., $O(2|\theta|)$), and self-attention networks for CITIES (i.e., $O(13d^2)$), which is higher than \proposed{}. For the time complexity, the naive combined model needs multiple steps 
composed of augmenting items in each user's sequence, training the CITIES model, and fine-tuning the sequence encoder, while \proposed{} is trained in an end-to-end manner, being 1.62 times faster for the model training. Our complexity analysis demonstrates the efficiency of \proposed{} in terms of both space and time.

\begin{table}[t]
\centering
\caption{Statistics of datasets. \textit{Int}. denotes interaction.}
\vspace{-2ex}
\resizebox{0.7\columnwidth}{!}{
\begin{tabular}{c|cccc}
\toprule
\textbf{Dataset}    & \textbf{\# Item} & \textbf{\# User} & \textbf{\# Int.} & \textbf{Avg $|\mathcal{S}_u|$} \\ \midrule \midrule
\textbf{Clothing} & 174,484 & 184,050 & 1,068,972 & 4.01 \\
\textbf{Sports} & 83,728 & 83,970 & 589,029 & 5.11\\
\textbf{Beauty} &  57,289 &52,204 & 394,908 & 5.6 \\
\textbf{Grocery}    & 39,264           & 32,126           & 275,256                 & 6.6                 \\
\textbf{Automotive} & 40,287           & 34,315           & 183,567                 & 3.6                 \\
\textbf{Music} & 20,356 & 20,165 & 132,595 & 5.11 \\

\textbf{Foursquare} & 13,335 & 43,110 & 306,553 & 5.12 \\
\textbf{Behance} & 32,491 & 28,915 & 712,271 & 22.7 \\
\bottomrule

\end{tabular}
}
\label{table:sta}
\vspace{-2ex}
\end{table}

\section{Experiments}

\subsection{Experimental Settings}
\label{exp:setting}
\subsubsection{Datasets. }
We compare~\proposed~with baseline models on \textit{eight} real-world datasets from the following online services: Amazon \cite{mcauley2015image}, Foursquare \cite{yuan2014graph}, and Behance \cite{he2016vista}.
Amazon datasets contain users' interactions with products, and we use six product categories from the Amazon datasets for evaluations. Foursquare and Behance datasets contain users' check-in data in different cities, and users' interactions with arts, respectively. For the data pre-processing, we filter out the users and items associated with fewer than 5 interactions to follow the previous studies \cite{li2020time,FMLP,caser,bert4rec}. In Table~\ref{table:sta}, we provide the statistics of datasets after preprocessing.

\begin{table*}[t]
\centering
\caption{Performance comparison. Improvements of~\proposed~vs. baseline models are measured in Overall and Mean (Head User, Tail User, Head Item, and Tail Item) performance based on average of HR@10 and ND@10. OOM: Out of memory on 16GB DGX.}
\vspace{-2.5ex}
\resizebox{0.82\textwidth}{!}{
\begin{tabular}{c|cl|cc|cc|cc|cc|cc|cc|cc}
\toprule
\multirow{2}{*}{\textbf{Dataset}} & \multicolumn{2}{c|}{\multirow{2}{*}{\textbf{Model}}} & \multicolumn{2}{c|}{\textbf{Overall}} & \multicolumn{2}{c|}{\textbf{Head User}} & \multicolumn{2}{c|}{\textbf{Tail User}} & \multicolumn{2}{c|}{\textbf{Head Item}} & \multicolumn{2}{c|}{\textbf{Tail Item}} & \multicolumn{2}{c|}{\textbf{Mean}}                     & \multicolumn{2}{c}{\textbf{Improvement (\%)}} \\
                                  & \multicolumn{2}{c|}{}               & HR@10             & ND@10             & HR@10              & ND@10              & HR@10              & ND@10              & HR@10              & ND@10              & HR@10              & ND@10              & \multicolumn{1}{c}{HR@10} & \multicolumn{1}{c|}{ND@10} & Overall                    & Mean              \\ \midrule

\multirow{7}{*}{Clothing}         & \multicolumn{2}{c|}{BERT4Rec}                        & 0.3892            & 0.2455            & 0.3852             & 0.2464             & 0.3903             & 0.2453             & \textbf{0.8097}    & 0.5145             & 0.0128             & 0.0048             & 0.3995                    & 0.2528                     & 11.4                     & 11.4               \\
                                  & \multicolumn{2}{c|}{Tail-Net}                        & 0.3769            & 0.2252            & 0.3692             & 0.2219             & 0.3789             & 0.2260             & 0.7882             & 0.4736             & 0.0087             & 0.0029             & 0.3863                    & 0.2311                     & 17.4                     & 17.7               \\
                                  & \multicolumn{2}{c|}{INSERT}                          & OOM               & OOM               & OOM                & OOM                & OOM                & OOM                & OOM                & OOM                & OOM                & OOM                & OOM     & OOM     & -                        & -                  \\
                                  & \multicolumn{2}{c|}{ASReP}                           & 0.4137            & 0.2550            & 0.4292             & 0.2698             & 0.4096             & 0.2511             & 0.7605             & 0.4894             & 0.1033             & 0.0452             & 0.4257                    & 0.2639                     & 5.7                      & 5.3                \\
                                  & \multicolumn{2}{c|}{SASRec}                          & 0.4054            & 0.2491            & 0.4202             & 0.2615             & 0.4015             & 0.2458             & 0.7476             & 0.4794             & 0.0991             & 0.0429             & 0.4171                    & 0.2574                     & 8.0                      & 7.7                \\
                                  & \multicolumn{2}{c|}{CITIES}                          & 0.4050            & 0.2474            & 0.4220             & 0.2601             & 0.4005             & 0.2440             & 0.7177             & 0.4621             & \textbf{0.1252}    & \textbf{0.0552}    & 0.4164                    & 0.2554                     & 8.3                      & 8.1                \\
                                  \rowcolor{gainsboro}\cellcolor{white} & \multicolumn{2}{c|}{MELT}        & \textbf{0.4350}   & \textbf{0.2718}   & \textbf{0.4462}    & \textbf{0.2803}    & \textbf{0.4321}    & \textbf{0.2696}    & 0.7864            & \textbf{0.5270}    & 0.1205             & 0.0435             & \textbf{0.4463}           & \textbf{0.2801}            & -                        & -                  \\ \midrule
\multirow{7}{*}{Sports}           & \multicolumn{2}{c|}{BERT4Rec}                        & 0.4886            & 0.3197            & 0.5402             & 0.3612             & 0.4782             & 0.3113             & 0.7988             & 0.5451             & 0.1627             & 0.0828             & 0.4950                    & 0.3251                     & 9.4                      & 8.9                \\
                                  & \multicolumn{2}{c|}{Tail-Net}                        & 0.4334            & 0.2718            & 0.4637             & 0.2975             & 0.4274             & 0.2667             & 0.8224             & 0.5227             & 0.0247             & 0.0083             & 0.4346                    & 0.2738                     & 25.4                     & 26.0               \\
                                  & \multicolumn{2}{c|}{INSERT}                          & 0.3987            & 0.2359            & 0.4167             & 0.2478             & 0.3950             & 0.2335             & 0.7403             & 0.4459             & 0.0397             & 0.0151             & 0.3979                    & 0.2356                     & 39.3                     & 40.9               \\
                                  & \multicolumn{2}{c|}{ASReP}                           & 0.5034            & 0.3249            & 0.5486             & 0.3603             & 0.4944             & 0.3178             & 0.8107             & 0.5498             & 0.1805             & 0.0886             & 0.5086                    & 0.3291                     & 6.7                      & 6.6                \\
                                  & \multicolumn{2}{c|}{SASRec}                          & 0.4972            & 0.3201            & 0.5439             & 0.3539             & 0.4878             & 0.3133             & 0.8110             & 0.5481             & 0.1676             & 0.0806             & 0.5026                    & 0.3240                     & 8.2                      & 8.0                \\
                                  & \multicolumn{2}{c|}{CITIES}                          & 0.5102            & 0.3283            & 0.5565             & 0.3603             & 0.5008             & 0.3219             & 0.7731             & 0.5328             & \textbf{0.2339}    & \textbf{0.1134}    & 0.5161                    & 0.3321                     & 5.4                      & 5.2                \\
                                  \rowcolor{gainsboro}\cellcolor{white} & \multicolumn{2}{c|}{MELT}        & \textbf{0.5377}   & \textbf{0.3463}   & \textbf{0.5848}    & \textbf{0.3780}    & \textbf{0.5282}    & \textbf{0.3399}    & \textbf{0.8429}    & \textbf{0.5943}    & 0.2169             & 0.0857             & \textbf{0.5432}           & \textbf{0.3495}            & -                        & -                  \\ \midrule
\multirow{7}{*}{Beauty}           & \multicolumn{2}{c|}{BERT4Rec}                        & 0.4476            & 0.3005            & 0.5083             & 0.3635             & 0.4337             & 0.2861             & 0.7777             & 0.5410             & 0.1087             & 0.0536             & 0.4571                    & 0.3111                     & 11.1                     & 10.8               \\
                                  & \multicolumn{2}{c|}{Tail-Net}                        & 0.4233            & 0.2643            & 0.4736             & 0.3180             & 0.4118             & 0.2520             & \textbf{0.7991}    & 0.5068             & 0.0375             & 0.0154             & 0.4305                    & 0.2731                     & 20.9                     & 21.0               \\
                                  & \multicolumn{2}{c|}{INSERT}                          & 0.4069            & 0.2447            & 0.4387             & 0.2724             & 0.3997             & 0.2383             & 0.7764             & 0.4730             & 0.0278             & 0.0103             & 0.4107                    & 0.2485                     & 27.6                     & 29.1               \\
                                  & \multicolumn{2}{c|}{ASReP}                           & 0.4680            & 0.3112            & 0.5402             & 0.3772             & 0.4515             & 0.2962             & 0.7530             & 0.5179             & 0.1755             & 0.0991             & 0.4801                    & 0.3226                     & 6.7                      & 6.0                \\
                                  & \multicolumn{2}{c|}{SASRec}                          & 0.4576            & 0.2995            & 0.5351             & 0.3687             & 0.4399             & 0.2837             & 0.7396             & 0.5013             & 0.1680             & 0.0924             & 0.4707                    & 0.3115                     & 9.8                      & 8.8                \\
                                  & \multicolumn{2}{c|}{CITIES}                          & 0.4599            & 0.3039            & 0.5447             & 0.3786             & 0.4406             & 0.2868             & 0.7063             & 0.4893             & 0.2071             & \textbf{0.1135}    & 0.4747                    & 0.3171                     & 8.8                      & 7.5                \\
                                  \rowcolor{gainsboro}\cellcolor{white} & \multicolumn{2}{c|}{MELT}        & \textbf{0.5012}   & \textbf{0.3300}   & \textbf{0.5673}    & \textbf{0.3837}    & \textbf{0.4861}    & \textbf{0.3178}    & 0.7806             & \textbf{0.5581}    & \textbf{0.2144}    & 0.0959             & \textbf{0.5121}           & \textbf{0.3389}            & -                        & -                  \\ \midrule
\multirow{7}{*}{Grocery}          & \multicolumn{2}{c|}{BERT4Rec}                        & 0.4590            & 0.3168            & 0.5506             & 0.4152             & 0.4380             & 0.2942             & 0.7872             & 0.5692             & 0.1166             & 0.0534             & 0.4731                    & 0.3330                     & 6.6                      & 6.1                \\
                                  & \multicolumn{2}{c|}{Tail-Net}                        & 0.4274            & 0.2812            & 0.5160             & 0.3772             & 0.4070             & 0.2591             & 0.7998             & 0.5360             & 0.0387             & 0.0152             & 0.4404                    & 0.2969                     & 16.7                     & 16.0               \\
                                  & \multicolumn{2}{c|}{INSERT}                          & 0.4236            & 0.2651            & 0.4878             & 0.3121             & 0.4088             & 0.2543             & \textbf{0.8165}    & 0.5147             & 0.0135             & 0.0047             & 0.4317                    & 0.2715                     & 20.1                     & 21.6               \\
                                  & \multicolumn{2}{c|}{ASReP}                           & 0.4622            & 0.3079            & 0.5575             & 0.4036             & 0.4403             & 0.2859             & 0.7544             & 0.5235             & 0.1573             & 0.0829             & 0.4774                    & 0.3240                     & 7.4                      & 6.7                \\
                                  & \multicolumn{2}{c|}{SASRec}                          & 0.4556            & 0.3036            & 0.5539             & 0.3957             & 0.4330             & 0.2824             & 0.7499             & 0.5206             & 0.1486             & 0.0771             & 0.4714                    & 0.3190                     & 8.9                      & 8.2                \\
                                  & \multicolumn{2}{c|}{CITIES}                          & 0.4588            & 0.3080            & 0.5593             & 0.4005             & 0.4357             & 0.2867             & 0.7245             & 0.5159             & 0.1815             & \textbf{0.0910}    & 0.4753                    & 0.3235                     & 7.9                      & 7.1                \\
                                  \rowcolor{gainsboro}\cellcolor{white} & \multicolumn{2}{c|}{MELT}        & \textbf{0.4910}   & \textbf{0.3360}   & \textbf{0.5802}    & \textbf{0.4235}    & \textbf{0.4705}    & \textbf{0.3159}    & 0.7805             & \textbf{0.5783}    & \textbf{0.1888}    & 0.0831             & \textbf{0.5050}           & \textbf{0.3502}            & -                        & -                  \\ \midrule
\multirow{7}{*}{Automotive}       & \multicolumn{2}{c|}{BERT4Rec}                        & 0.3495            & 0.2093            & 0.3762             & 0.2240             & 0.3433             & 0.2059             & \textbf{0.8055}    & 0.4839             & 0.0041             & 0.0013             & 0.3823                    & 0.2288                     & 17.9                     & 15.7               \\
                                  & \multicolumn{2}{c|}{Tail-Net}                        & 0.3521            & 0.2118            & 0.3819             & 0.2322             & 0.3453             & 0.2071             & 0.7893             & 0.4818             & 0.0211             & 0.0074             & 0.3844                    & 0.2321                     & 16.9                     & 14.6               \\
                                  & \multicolumn{2}{c|}{INSERT}                          & 0.3516            & 0.2113            & 0.3796             & 0.2296             & 0.3451             & 0.2070             & 0.7886             & 0.4807             & 0.0206             & 0.0072             & 0.3835                    & 0.2311                     & 17.1                     & 15.0               \\
                                  & \multicolumn{2}{c|}{ASReP}                           & 0.3603            & 0.2229            & 0.4084             & 0.2586             & 0.3492             & 0.2147             & 0.6780             & 0.4342             & 0.1197             & 0.0629             & 0.3888                    & 0.2426                     & 13.0                     & 11.9               \\
                                  & \multicolumn{2}{c|}{SASRec}                          & 0.3472            & 0.2154            & 0.3911             & 0.2464             & 0.3371             & 0.2083             & 0.6570             & 0.4229             & 0.1126             & 0.0583             & 0.3745                    & 0.2340                     & 17.1                     & 16.2               \\
                                  & \multicolumn{2}{c|}{CITIES}                          & 0.3505            & 0.2222            & 0.3927             & 0.2518             & 0.3408             & 0.2153             & 0.6348             & 0.4242             & 0.1352             & \textbf{0.0692}    & 0.3759                    & 0.2401                     & 15.1                     & 14.7               \\
                                  \rowcolor{gainsboro}\cellcolor{white} & \multicolumn{2}{c|}{MELT}        & \textbf{0.4024}   & \textbf{0.2566}   & \textbf{0.4397}    & \textbf{0.2850}    & \textbf{0.3938}    & \textbf{0.2501}    & 0.7284             & \textbf{0.5098}    & \textbf{0.1556}    & 0.0649             & \textbf{0.4294}           & \textbf{0.2775}            & -                        & -                  \\ \midrule

\multirow{7}{*}{Music}            & \multicolumn{2}{c|}{BERT4Rec}                        & 0.5203            & 0.3628            & 0.6443             & 0.4567             & 0.4835             & 0.3349             & 0.8448             & 0.6335             & 0.2270             & 0.1181             & 0.5499                    & 0.3858                     & 13.9                     & 12.5               \\
                                  & \multicolumn{2}{c|}{Tail-Net}                        & 0.4791            & 0.3072            & 0.5812             & 0.3746             & 0.4487             & 0.2871             & 0.8729             & 0.5921             & 0.1231             & 0.0496             & 0.5065                    & 0.3259                     & 27.9                     & 26.4               \\
                                  & \multicolumn{2}{c|}{INSERT}                          & 0.4400            & 0.2724            & 0.5077             & 0.3103             & 0.4198             & 0.2611             & 0.8346             & 0.5384             & 0.0832             & 0.0319             & 0.4613                    & 0.2854                     & 41.1                     & 40.9               \\
                                  & \multicolumn{2}{c|}{ASReP}                           & 0.5443            & 0.3830            & 0.6677             & 0.4724             & 0.5076             & 0.3564             & 0.8547             & 0.6341             & 0.2637             & 0.1559             & 0.5734                    & 0.4047                     & 8.4                      & 7.6                \\
                                  & \multicolumn{2}{c|}{SASRec}                          & 0.5344            & 0.3717            & 0.6602             & 0.4643             & 0.4970             & 0.3442             & 0.8512             & 0.6269             & 0.2481             & 0.1410             & 0.5641                    & 0.3941                     & 11.0                     & 9.8                \\
                                  & \multicolumn{2}{c|}{CITIES}                          & 0.5537            & 0.3879            & 0.6850             & 0.4841             & 0.5147             & 0.3593             & 0.8310             & 0.6428             & 0.3030             & \textbf{0.1574}    & 0.5834                    & 0.4109                     & 6.8                      & 5.8                \\
                                  \rowcolor{gainsboro}\cellcolor{white} & \multicolumn{2}{c|}{MELT}        & \textbf{0.5997}   & \textbf{0.4058}   & \textbf{0.7091}    & \textbf{0.4846}    & \textbf{0.5672}    & \textbf{0.3824}    & \textbf{0.8961}    & \textbf{0.6835}    & \textbf{0.3318}    & 0.1548             & \textbf{0.6261}           & \textbf{0.4263}            & -                        & -                  \\ \midrule

\multirow{7}{*}{Foursquare}       & \multicolumn{2}{c|}{BERT4Rec}                        & 0.8939            & 0.8058            & 0.9198             & 0.8406             & 0.8756             & 0.7813             & 0.9445             & 0.8551             & 0.1670             & 0.0975             & 0.7267                    & 0.6436                     & 2.8                      & 8.6                \\
                                  & \multicolumn{2}{c|}{Tail-Net}                        & 0.8768            & 0.7828            & 0.9081             & 0.8210             & 0.8548             & 0.7560             & 0.9339             & 0.8358             & 0.0581             & 0.0226             & 0.6887                    & 0.6089                     & 5.3                      & 14.7               \\
                                  & \multicolumn{2}{c|}{INSERT}                          & 0.8490            & 0.7215            & 0.8998             & 0.7784             & 0.8398             & 0.7112             & 0.9326             & 0.8051             & 0.2677             & 0.1400             & 0.7350                    & 0.6087                     & 11.3                     & 10.7               \\
                                  & \multicolumn{2}{c|}{ASReP}                           & 0.9243            & 0.8189            & 0.9444             & 0.8527             & 0.9102             & 0.7952             & \textbf{0.9656}    & 0.8616             & 0.3314             & 0.2066             & 0.7879                    & 0.6790                     & 0.3                      & 1.4                \\
                                  & \multicolumn{2}{c|}{SASRec}                          & 0.9193            & 0.8124            & 0.9409             & 0.8474             & 0.9040             & 0.7878             & 0.9624             & 0.8556             & 0.2996             & 0.1918             & 0.7767                    & 0.6707                     & 0.9                      & 2.8                \\
                                  & \multicolumn{2}{c|}{CITIES}                          & 0.9209            & 0.8191            & 0.9427             & 0.8540             & 0.9056             & 0.7944             & 0.9608             & 0.8609             & 0.3474             & \textbf{0.2180}    & 0.7891                    & \textbf{0.6818}            & 0.5                      & 1.1                \\
                                  \rowcolor{gainsboro}\cellcolor{white} & \multicolumn{2}{c|}{MELT}        & \textbf{0.9271}   & \textbf{0.8210}   & \textbf{0.9471}    & \textbf{0.8541}    & \textbf{0.9131}    & \textbf{0.7977}    & 0.9628             & \textbf{0.8644}    & \textbf{0.4156}    & 0.1966             & \textbf{0.8097}           & 0.6782                     & \textbf{-}               & -                  \\ \midrule
\multirow{7}{*}{Behance}          & \multicolumn{2}{c|}{BERT4Rec}                        & 0.6359                 & 0.4388                 & 0.6490                  & 0.4576                  &  0.6225                 & 0.4197                  &  0.7856                 & 0.5447                  & 0.0328                  & 0.0123                  & 0.5225     & 0.3586     & 20.3                        & 32.3                   \\
                                  & \multicolumn{2}{c|}{Tail-Net}                        & 0.6039            & 0.4241            & 0.6189             & 0.4450             & 0.5885             & 0.4027             & 0.7452             & 0.5259             & 0.0344             & 0.0137             & 0.4968                    & 0.3468                     & 25.8                     & 38.2               \\
                                  & \multicolumn{2}{c|}{INSERT}                          & 0.5613            & 0.3454            & 0.5542             & 0.3381             & 0.5685             & 0.3529             & 0.6998             & 0.4308             & 0.0032             & 0.0013             & 0.4564                    & 0.2808                     & 42.6                     & 58.1               \\
                                  & \multicolumn{2}{c|}{ASReP}                           & 0.7163            & 0.5177            & 0.7411             & 0.5416             & 0.6910              & 0.4934             & 0.8147             & 0.5964             & 0.3197             & 0.2006    & 0.6416                    & 0.4580                     & 4.8                      & 6.0                \\
                                  & \multicolumn{2}{c|}{SASRec}                          & 0.6891            & 0.4908            & 0.7158             & 0.5146             & 0.6618             & 0.4666             & 0.7912             & 0.5748             & 0.2778             & 0.1526             & 0.6117                    & 0.4272                     & 9.6                      & 12.2               \\
                                  & \multicolumn{2}{c|}{CITIES}                          & 0.7107            & 0.5156            & 0.7370              & 0.5386             & 0.6839             & 0.4921             & 0.8030              & 0.5938             & 0.3386             & 0.2003             & 0.6406                    & 0.4562                     & 5.4                      & 6.3                \\
                                  \rowcolor{gainsboro}\cellcolor{white} & \multicolumn{2}{c|}{MELT}        & \textbf{0.7505}   & \textbf{0.5424}   & \textbf{0.7736}    & \textbf{0.5642}    & \textbf{0.7268}    & \textbf{0.5201}    & \textbf{0.8336}    & \textbf{0.6267}    & \textbf{0.4154}    & \textbf{0.2025}             & \textbf{0.6874}           & \textbf{0.4784}            & -                        & -                  \\ \bottomrule
\end{tabular}

}

\label{tab:main}
\vspace{-1ex}
\end{table*}

\subsubsection{Evaluation Protocol. }
Following the leave-one-out protocol \cite{sasrec,bert4rec}, we use the most recent interaction in each user's sequence for testing, the second most recent interaction for validation, and the remaining interactions for training. 
We measure the recommendation performance with Hit Ratio (\textbf{HR@K}) and Normalized Discounted Cumulative Gain (\textbf{ND@K}) \cite{INSERT,bert4rec}. 
To reduce the computational complexity in the evaluation, we pair each user's ground-truth item with 100 randomly sampled items not interacted by the user \cite{FMLP,sasrec}. We report the performance of head users, tail users, head items, and tail items along with the overall performance, where head and tail ones are separated by $\alpha=20\%$ for Amazon datasets and $50\%$ for others (Section \ref{sec:ht_split})\cite{CITIES,chen2020esam,TailNet}.
Additionally, we provide the average performance (\textbf{Mean}) of head user, tail user, head item, and tail item.
We train every compared model five times, and report the average of the metrics \cite{chen2020esam,hyun2020interest,hyun2022beyond}.

\vspace{-1.0ex}
\subsubsection{Compared Methods.}
We compare our proposed method with the following baseline methods. \textbf{Standard SRS}: 1) SASRec \cite{sasrec} uses an unidirectional transformer to predict the next item. 2) BERT4Rec \cite{bert4rec} utilizes a bidirectional transformer through the Cloze objective losses. 3) FMLP \cite{FMLP} replaces the self-attention with a simple MLP, and reduces noise from item representations by introducing 1-D Fourier Transform. \textbf{SRS for long-tailed users}: 4) ASReP \cite{ASReP} explicitly augments tail users' interactions by reversely training the SRS model. 
5) INSERT \cite{INSERT} is a session-based model that utilizes the preference of users who are similar to a target user. 
\textbf{SRS for long-tailed items}: 6) CITIES \cite{CITIES} adopts the self-attention mechanism to utilize head items to enhance the representation of tail items.
7) Tail-Net \cite{TailNet} explicitly weights items according to the ratio of head and tail items in each user's sequence in the inference time.

\begin{figure*}[t]
    \begin{center}
        \includegraphics[width=0.87\linewidth]{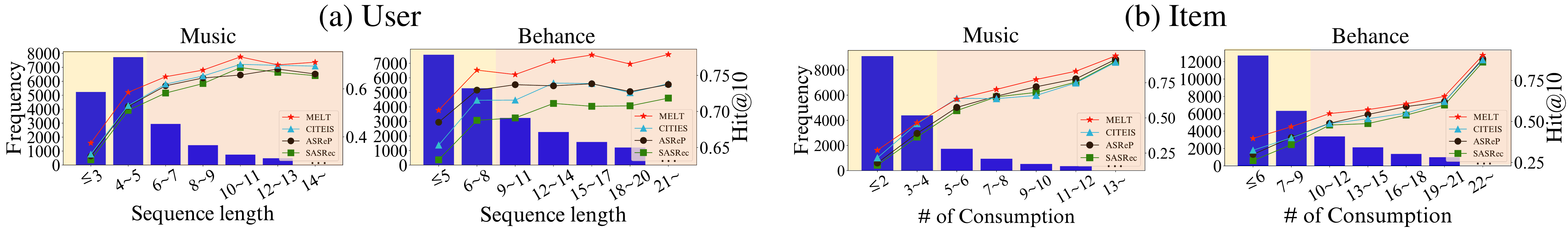}
    \end{center}
    \vspace{-3.5ex}
    \caption{Performance comparison with sequence length and the number of consumptions in terms of users and items, respectively. Red and Yellow backgrounds denote the head and tail categories, respectively.}
    \label{fig:per_seq}
    \vspace{-1ex}
\end{figure*}

\subsubsection{Implementation Details.}
\label{sec:imp_detail}
To evaluate the performance of each method, we first find the optimal hyperparameters under a single seed based on the validation set, and then train them with 5 different random seeds given the optimal hyperparameters found.

\smallskip
\noindent\textbf{Compared Methods. }
For fair comparisons, we set the hidden dimension size (i.e., $d$) to 50 and maximum sequence length to 50 \cite{sasrec,ASReP,FMLP}. We add padded items if the sequence length is lower than 50 and use the recent 50 interactions if the sequence length is longer than 50. The batch size of self-attention-based methods (i.e., ASReP, SASRec, CITIES, and BERT4Rec), and others is set to 128 and 256, respectively.
We set the number of heads to 2 for the self-attention-based methods as it yields the best performance. 
Note that to prevent any information leakage, we exclude the user-item interaction in validation and test data when 1) reversely training the ASReP, and 2) estimating similar users to complement their short sequence in INSERT, both of which have not been considered in the authors' implementations.

\smallskip
\noindent \textbf{Our Proposed Framework. }
In order to find the optimal hyperparameters of \proposed{}, we adopt grid search in the range of $\{0.1,0.2,0.3,0.4\}$
for $\lambda_{\mathcal{U}}$ and $\lambda_{\mathcal{I}}$.
To reduce the complexity of the hyperparameter search, we set $\beta$ and $\gamma$ to 1 and 0, respectively.
In Section~\ref{sec:bbu} and~\ref{sec:bbi}, we implement the embedding generators ${G}_\phi^{\mathcal{U}}$ and ${G}_\phi^{\mathcal{I}}$ by a single-layer feed forward neural network without non-linearity to reduce the model complexity.
 $R$ and $K$ in Section~\ref{sec:bbu} and Section~\ref{sec:bbi} are randomly sampled per user and item from $\{1,2,3,...,\kappa_{\mathcal{U}} \}$ and $\{1,2,3,...,\kappa_{\mathcal{I}}\}$, respectively, where $\kappa_{\mathcal{U}}$ and $\kappa_{\mathcal{I}}$ are the number of interactions associated with users and items to distinguish the head and tail group. 
 More precisely, users whose sequences are longer than $\kappa_{\mathcal{U}}$ are head users, and items with the number of interactions more than $\kappa_{\mathcal{I}}$ are head items.
 
To ensure a complete representation of head items (i.e., $q_{i^H}$), and reduce noise in the mutual enhancement step, we firstly adopt a pretrained sequence encoder (i.e., $f_{\theta}(\cdot)$) that produces the user and item representation (i.e., $p_u$ and $q_i$), and fine-tune $\theta$ and $q_i$ during training in Section~\ref{sec:model_train}.
To construct $\mathcal{C}_i$ for item $i$ in the item branch, we also consider the reverse direction of the sequence for fully utilizing the interacted information, which starts with item $i$. In the example of Section~\ref{sec:bbi}, a subsequence $[i_4, i_2]$ is added to $\mathcal{C}_2$.

\looseness=-1
Since sequential models (e.g., SASRec \cite{sasrec} and FMLP \cite{FMLP}) used in this work do not explicitly assign explicit user representations (i.e., $p_u$), we use $p_u=f_{\theta}(\mathcal{S}_u)$ instead of assigning explicit user representations. 
Note that we explicitly assign a trainable embedding vector $q_i$ to each item $i$.

\begin{table}[t]
\centering
\caption{Performance based on FMLP \cite{FMLP} sequence encoder.}
\vspace{-2ex}
\resizebox{0.98\linewidth}{!}{
\begin{tabular}{c|c|cc|cc|cc}
\toprule
\multirow{2}{*}{\textbf{Data}} & \multirow{2}{*}{\textbf{Model}} & \multicolumn{2}{c|}{\textbf{Overall}} & \multicolumn{2}{c|}{\textbf{Tail User}} & \multicolumn{2}{c}{\textbf{Tail Item}} \\ \cline{3-8} 
                               &                                 & HR@10             & ND@10              & HR@10              & ND@10               & HR@10              & ND@10              \\ \midrule
\multirow{3}{*}{Clothing}      & FMLP                            & 0.392            & 0.242             & 0.389              & 0.239              & 0.127              & 0.062             \\
                               & +CITIES                         & 0.402             & 0.249             & 0.396              & 0.245              & 0.158              & 0.078             \\
                               & +MELT                           & \textbf{0.434}    & \textbf{0.274}    & \textbf{0.426}     & \textbf{0.269}     & \textbf{0.206}     & \textbf{0.091}    \\ \midrule

\multirow{3}{*}{Sports}      & FMLP                            & 0.513            & 0.337             & 0.502              & 0.329              & 0.220              & 0.118             \\
                               & +CITIES                         & 0.523             & 0.343             & 0.512              & 0.335              & 0.239              & \textbf{0.128}             \\
                               & +MELT                           & \textbf{0.542}    & \textbf{0.356}    & \textbf{0.531}     & \textbf{0.349}     & \textbf{0.274}     & 0.126    \\ \midrule

\multirow{3}{*}{Beauty}        & FMLP                            & 0.474             & 0.323             & 0.458              & 0.307              & 0.204              & 0.124             \\
                               & +CITIES                         & 0.491             & 0.334             & 0.475              & 0.319              & 0.224              & \textbf{0.135}    \\
                               & +MELT                           & \textbf{0.500}        & \textbf{0.341}        & \textbf{0.483}         & \textbf{0.326}         & \textbf{0.241}         & 0.127                 \\ \midrule
\multirow{3}{*}{Grocery}      & FMLP                            & 0.462            & 0.314             & 0.441              & 0.293              & 0.169              & 0.095             \\
                               & +CITIES                         & 0.479             & 0.328             & 0.459              & 0.308              & 0.188              & \textbf{0.105}             \\
                               & +MELT                           & \textbf{0.488}    & \textbf{0.333}    & \textbf{0.467}     & \textbf{0.313}     & \textbf{0.206}     & 0.097    \\ \midrule 

\multirow{3}{*}{Automotive}      & FMLP                            & 0.355            & 0.228             & 0.337              & 0.208              & 0.135              & 0.075             \\
                               & +CITIES                         & 0.384             & 0.249             & 0.375              & 0.243              & 0.149              & 0.084             \\
                               & +MELT                           & \textbf{0.390}    & \textbf{0.253}    & \textbf{0.380}     & \textbf{0.246}     & \textbf{0.174}     & \textbf{0.086}    \\ \midrule
\multirow{3}{*}{Music}         & FMLP                            & 0.553             & 0.394             & 0.518              & 0.366              & 0.275              & 0.141             \\
                               & +CITIES                         & 0.567             & 0.406    & 0.531              & 0.378              & 0.301              & \textbf{0.177}    \\
                               & +MELT                           & \textbf{0.590}    & \textbf{0.407}             & \textbf{0.555}     & \textbf{0.379}     & \textbf{0.341}     & 0.173             \\ \midrule

\multirow{3}{*}{Foursquare}      & FMLP                            & 0.917            & 0.773             & 0.901              & 0.743              & 0.350              & \textbf{0.241}             \\
                               & +CITIES                         & 0.920             & 0.820             & 0.904              & 0.795              & 0.338              & 0.224             \\
                               & +MELT                           & \textbf{0.925}    & \textbf{0.821}    & \textbf{0.910}     & \textbf{0.797}     & \textbf{0.409}     & 0.197    \\ \midrule
\multirow{3}{*}{Behance}       & FMLP                            & 0.729             & 0.531             & 0.704              & 0.506              & 0.394              & 0.272             \\
                               & +CITIES                         & 0.736             & 0.539             & 0.713              & 0.516              & 0.435              & \textbf{0.285}    \\
                               & +MELT                           & \textbf{0.751}    & \textbf{0.549}    & \textbf{0.731}     & \textbf{0.529}     & \textbf{0.474}     & 0.273             \\ \bottomrule

\end{tabular}
}
\label{tab:fmlp}
\vspace{-1ex}
\end{table}

\subsection{Overall Performance Comparison}
\label{sec:overall}
\looseness=-1
Table~\ref{tab:main} shows the performance of \proposed{} and baseline models. As~\proposed~is a model-agnostic framework, we also apply our framework to the state-of-the-art SRS (i.e., FMLP) in Table~\ref{tab:fmlp}.
We have the following observations: 1)~\proposed~significantly improves the overall performance on all datasets, even in the FMLP sequence encoder, which verifies the effectiveness of \proposed{}. 2)~\proposed~shows large improvements on both head and tail user groups compared with the baseline methods designed specifically for long-tailed users, i.e., ASReP and INSERT. It is clearly observed in Figure~\ref{fig:per_seq}(a) that \proposed{} outperforms the baselines over all sequence lengths. This implies that the knowledge in the user branch is helpful for supplementing the representation of tail users, while head and tail users can benefit from the enhanced representations of tail items obtained from the item branch.
3)~\proposed~improves the performance on the tail item group without sacrificing the performance on the head item group. Meanwhile, for most of the datasets, we find that CITIES sacrifices the performance on the head item group to improve the performance on the tail item group. This observation is clearly shown in Figure~\ref{fig:per_seq}(b) where \proposed{} shows the superiority on extreme long-tailedness (i.e., $\leq 2$ on Music and $\leq 6$ on Behance) while not sacrificing the performance on the head group.
Moreover, since CITIES ignores the long-tailed user problem, they perform poorly on the tail user group.
To sum up, we argue that the long-tailed user and item problems arise together in real-world datasets, and that they should be jointly addressed as proposed in \proposed{}.

\subsection{Fine-grained Performance Comparison}
\looseness=-1
\label{sec:fine_grained}
In Table~\ref{tab:joint},
we further investigate the benefit of \proposed{} by considering fine-grained scenarios of recommendation, i.e., recommending head/tail items to head/tail users. We use SASRec as the backbone model, and have the following observations.
1) \proposed{} generally surpasses ASReP, which considers only the long-tailed users, specifically in the scenario of recommending \textit{tail items} to users (i.e., HU/TI and TU/TI) thanks to alleviating the long-tailed item problem. 
2) \proposed{} is superior compared with CITIES, which addresses only the long-tailed item problem, in the scenario of recommending head items to \textit{tail users} (i.e., TU/HI) thanks to addressing the long-tailed user problem. 
3) \proposed{} outperforms other baselines, especially ASReP+CITIES, under the extreme scenario (i.e., TU/TI), which implies that a naive combination of two existing baselines is non-trivial and it is important to encourage the models to mutually enhanced each other in an end-to-end manner. 
4) SASRec, which overlooks the long-tailed problems, produces lower performance compared with other models devised to address the long-tailed problems. To sum up, \proposed{} is effective under different scenarios of recommendation without sacrificing the performance of specific scenarios (i.e., Mean) compared with the baseline models that only consider either the long-tailed user or item problem.

\begin{table}[t]
\centering
\caption{Fine-grained performance comparison (HR@10)
($\textbf{H}$: Head,
$\textbf{T}$: Tail,
$\textbf{U}$: User,
$\textbf{I}$: Item), e.g., $\textbf{HU/TI}$ is recommending tail items to head users. Mean: the mean of four scores.}
\vspace{-2ex}
\resizebox{0.95\columnwidth}{!}{
\begin{tabular}{c|c|cccc|c}
\toprule
\textbf{\textbf{\textbf{Data}}} & \textbf{\textbf{\textbf{Model}}} & \textbf{\textbf{HU/HI}} & \textbf{\textbf{HU/TI}} & \textbf{\textbf{TU/HI}} & \textbf{\textbf{TU/TI}} & \textbf{\textbf{\textbf{Mean}}} \\ \midrule
\multirow{5}{*}{Music}          & SASRec                           & 0.912                   & 0.325                   & 0.828                   & 0.230                   & 0.574                           \\
                                & ASReP                            & 0.915                   & 0.339                   & 0.832                   & 0.246                   & 0.583                           \\
                                & CITIES                           & 0.896                   & 0.405                   & 0.807                   & 0.279                   & 0.597                           \\
                                & ASReP+CITIES                     & 0.915                   & 0.368                   & 0.828                   & 0.273                   & 0.596                           \\
                                & \textbf{\proposed}                    & \textbf{0.928}          & \textbf{0.418}          & \textbf{0.884}          & \textbf{0.312}          & \textbf{0.636}                  \\ \midrule
\multirow{5}{*}{Beauty}         & SASRec                           & 0.807                   & 0.240                   & 0.724                   & 0.152                   & 0.481                           \\
                                & ASReP                            & 0.811                   & 0.245                   & 0.739                   & 0.160                   & 0.489                           \\
                                & CITIES                           & 0.786                   & 0.282                   & 0.687                   & 0.191                   & 0.487                           \\
                                & ASReP+CITIES                     & \textbf{0.820}          & 0.245                   & 0.743                   & 0.169                   & 0.494                           \\
                                & \textbf{\proposed}                    & 0.818                   & \textbf{0.291}          & \textbf{0.775}          & \textbf{0.197}          & \textbf{0.520}                  \\ \midrule
\multirow{5}{*}{Automotive}         & SASRec                           & 0.701                 & 0.133                   & 0.646                   & 0.108                   & 0.397                           \\
                                & ASReP                            & 0.724                   & 0.146                   & 0.667                   & 0.114                   & 0.413                           \\
                                & CITIES                           & 0.675                   & 0.158                   & 0.625                   & 0.130                   & 0.397                           \\
                                & ASReP+CITIES                     & 0.726          & 0.147                  & 0.668                   & 0.124                   & 0.416                           \\
                                & \textbf{\proposed}                    & \textbf{0.748}                   & \textbf{0.184}          & \textbf{0.724}          & \textbf{0.149}          & \textbf{0.451}                  \\ \midrule
\multirow{5}{*}{Behance}        & SASRec                           & 0.816                   & 0.320                   & 0.766                   & 0.234                   & 0.534                           \\
                                & ASReP                            & 0.840                   & 0.351                   & 0.789                   & 0.287                   & 0.567                           \\
                                & CITIES                           & 0.828                   & 0.377                   & 0.778                   & 0.298                   & 0.570                           \\
                                & ASReP+CITIES                     & 0.839                   & 0.349                   & 0.792                   & 0.296                   & 0.569                           \\
                                & \textbf{\proposed}                    & \textbf{0.854}          & \textbf{0.458}          & \textbf{0.813}          & \textbf{0.371}          & \textbf{0.624}                  \\ \bottomrule
\end{tabular}
}
\label{tab:joint}
\end{table}

\begin{figure}[t]
\vspace{-2ex}
    \begin{center}
        \includegraphics[width=0.99\columnwidth]{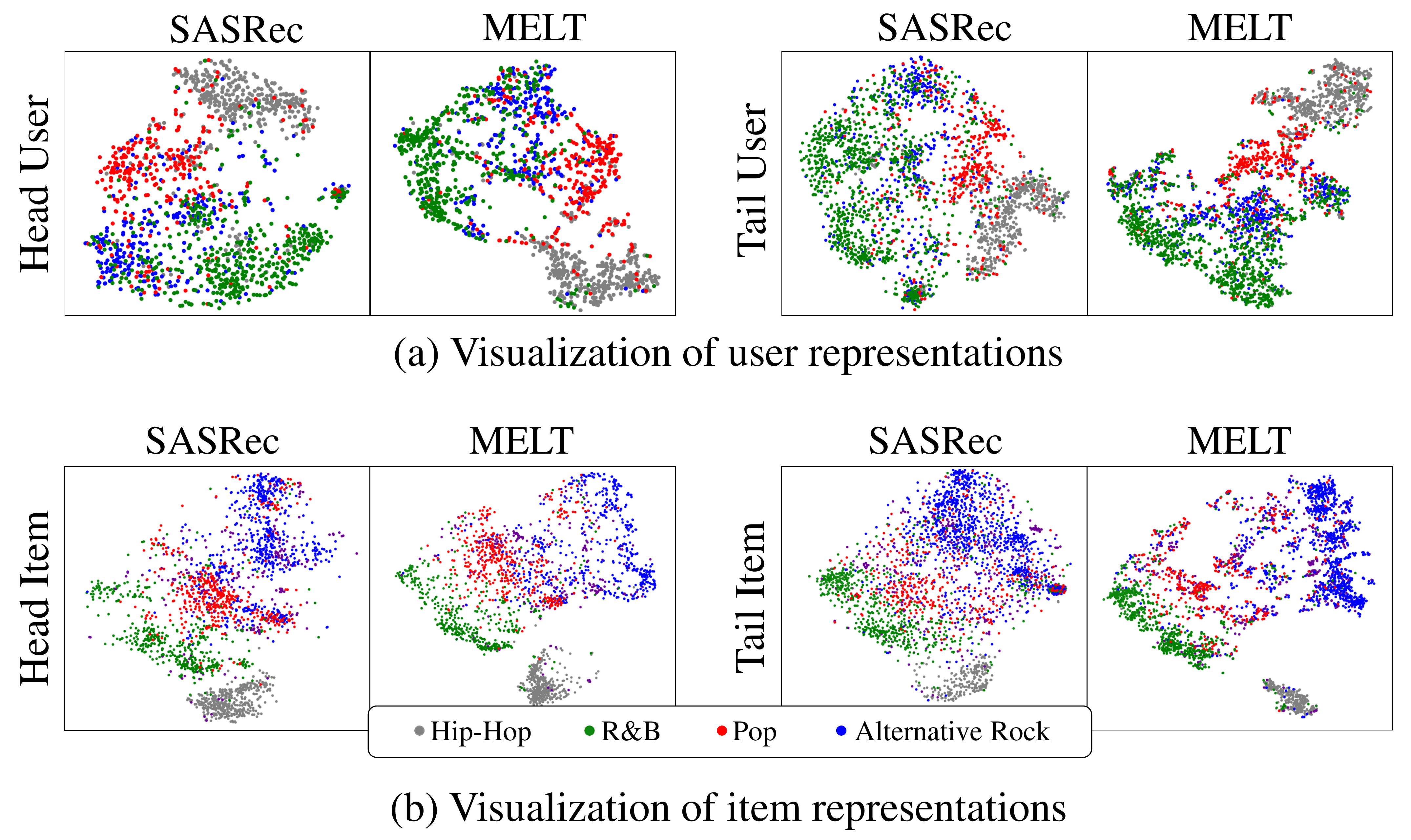} 
    \end{center}
    \vspace{-2ex}
    \caption{Visualization of user and item representations in Amazon Music dataset.}
    \label{fig:visual}
    \vspace{-2ex}
\end{figure}

\subsection{Analysis on User and Item Representations}
\looseness=-1
To study the enhancement of tail user and tail item representations, we visualize the users' representations (Figure~\ref{fig:visual}(a)) and items' representations (Figure~\ref{fig:visual}(b)) on Amazon Music data by using t-SNE. In Figure~\ref{fig:visual}(a), 
we color each user's representation according to the test item's category assuming that similar users consume the same category. 
We observe that for head users, thanks to the mutual enhancement of bilateral branches where the representation of tail items in a user's sequence is being enhanced, similar users are more closely grouped in \proposed{} than in SASRec. Besides, \proposed{} is also superior in clustering the tail users, which indicates that \proposed{} enhances the tail users' representations via not only enhanced representation of tail items in a user sequence, but also from the knowledge transferred from the head users.
In Figure~\ref{fig:visual}(b), we visualize the head and tail item representations according to the item category. For head items, they are well-clustered in both SASRec and \proposed{}. On the other hand, \proposed~is superior to SASRec in clustering the tail items according to their category, which indicates that \proposed{} indeed enhances the tail item representations.

\begin{table}[t]
\centering
\caption{Ablation studies (HR@10) ($\text{\textbf{U}}$: User branch, 
$\text{\textbf{I}}$: Item branch, 
$\text{\textbf{M}}$: Mutual enhancement,
$\text{\textbf{C}}$: Curriculum learning)
}
\vspace{-1ex}
\resizebox{0.95\columnwidth}{!}{
\begin{tabular}{c|c|cccc|ccccc|c}
\toprule
\multirow{1}{*}{\textbf{Data}}& \multirow{1}{*}{\textbf{Row}}   &\multirow{1}{*}{\textbf{U}} & \multirow{1}{*}{\textbf{I}} & \multirow{1}{*}{\textbf{M}}             & \multirow{1}{*}{\textbf{C}} & 

                     \textbf{Over.} & \textbf{HU}   & \textbf{TU}   & \textbf{HI}   & \textbf{TI}   & \textbf{Mean}  \\ \midrule
\multirow{6}{*}{\textbf{Music}} &  1 &                          &                             &                                             &                              &       0.534           &   0.660             &        0.497        &      0.851          &    0.248            &       0.564         \\
                                  & 2 &\color{black}\checkmark     & \multicolumn{1}{l}{}        &                                             & \multicolumn{1}{l|}{}        &         0.541         &      0.659          &      0.506          &  0.891              &     0.225           &      0.570          \\
                                           &    3  &             & \color{black}\checkmark     &                                             &                              &   0.578               &        0.698        &     0.543           &   0.848             &       0.331         &      0.605          \\
                                  & 4 &\color{black}\checkmark     & \color{black}\checkmark     &                                             &                              &  0.582                &      0.699          &       0.548         &    0.877            &      0.316          &   0.610             \\
                                  & 5 &\color{black}\checkmark     & \color{black}\checkmark     & \multicolumn{1}{c}{\color{black}\checkmark} & \multicolumn{1}{l|}{}        &      0.597            &     \textbf{0.711}           &      0.563          &      0.891          &     0.330           &       0.623         \\
                                  & 6 &\color{black}\checkmark     & \color{black}\checkmark     & \multicolumn{1}{c}{\color{black}\checkmark} & \color{black}\checkmark      &        \textbf{0.600}         &      0.709          &      \textbf{0.567}          &       \textbf{0.896}         &     \textbf{0.332}           &    \textbf{0.626}            \\ \midrule
\multirow{6}{*}{\textbf{Beauty}}  &     1  &                      &                             &                                             &                              & 0.458            & 0.540          & 0.452          & 0.753          & 0.177          & 0.480          \\
                                   & 2&\color{black}\checkmark     & \multicolumn{1}{l}{}        &                                             & \multicolumn{1}{l|}{}        & 0.485            & 0.545          & 0.472          & \textbf{0.806} & 0.156          & 0.495          \\
                                  &     3         &          & \color{black}\checkmark     &                                             &                              & 0.471            & 0.540          & 0.459          & 0.724          & 0.212          & 0.484          \\
                                  & 4 &\color{black}\checkmark     & \color{black}\checkmark     &                                             &                              & 0.490            & 0.550          & 0.476          & 0.777          & 0.195          & 0.499          \\
                                  & 5 &\color{black}\checkmark     & \color{black}\checkmark     & \multicolumn{1}{c}{\color{black}\checkmark} &                              & 0.498            & 0.565          & 0.483          & 0.775          & \textbf{0.213} & 0.509          \\
                                  & 6 &\color{black}\checkmark     & \color{black}\checkmark     & \multicolumn{1}{c}{\color{black}\checkmark} & \color{black}\checkmark      & \textbf{0.502}   & \textbf{0.566} & \textbf{0.487} & 0.783          & \textbf{0.213}          & \textbf{0.512}  \\ \bottomrule
\end{tabular}
}

\vspace{-1ex}
\label{tab:ablation}
\end{table}

\subsection{Ablation Studies}
\label{sec:ablation}
In Table~\ref{tab:ablation}, we conduct ablation studies to understand the benefit of each component of~\proposed. For ablation studies, we use SASRec as the backbone.
The variant that does not use any of the components (row 1 in each dataset) is equivalent to SASRec. We have the following observations: 
1) \textbf{Effect of each branch}: The user (row 2) and item (row 3) branches show improvements on tail users and items, respectively. This demonstrates that the representation of tail users and items are enhanced through each branch, respectively. 
2) \textbf{Effect of combining bilateral branches} (row 4):
Despite its simplicity upon considering user and item branches together (i.e., by excluding Equation~\ref{eqn:eq12} meaning that the knowledge is transferred from the item branch to user branch only\footnote{Note that the ablation study of knowledge transferred from the item branch to the user branch via $G^\mathcal{I}_{\phi}$ cannot be conducted since the item branch directly updates the item embeddings.}),
the overall performance increases thanks to addressing both the long-tailed user and item problems. However, even with the help of the item branch, the performance of tail items deteriorates as the user branch enhances the recommendation performance on tail users by largely recommending head items (compare \textbf{HI} scores between rows 1 and 2). Thus, simply combining the user and item branches suffers from the negative impact between the branches.
3) \textbf{Effect of the mutual enhancement} (row 5): 
The mutual enhancement between the bilateral branches further improves the overall performance by making both branches enhance each other. Specifically, the mutual enhancement mechanism alleviates the negative impact between the user and item branches (i.e., degraded performance of tail items in row 4), resulting in the enhanced performance on tail items. We speculate that the knowledge obtained from the user branch enhances the item branch (Equation \ref{eqn:eq12}) so that the performance of tail items can be further enhanced. 
This corroborates our argument that considering the counterpart model is crucial for jointly addressing the long-tailed user and item problems, and \proposed{} achieves this by iteratively enhancing each other in an end-to-end manner. 
4) \textbf{Effect of curriculum learning} (row 6): The proposed CL strategy improves the overall performance thanks to the stabilization in the learning process.

\begin{figure}[t]
    \vspace{-1ex}
    \begin{center}
        \includegraphics[width=0.80\columnwidth]{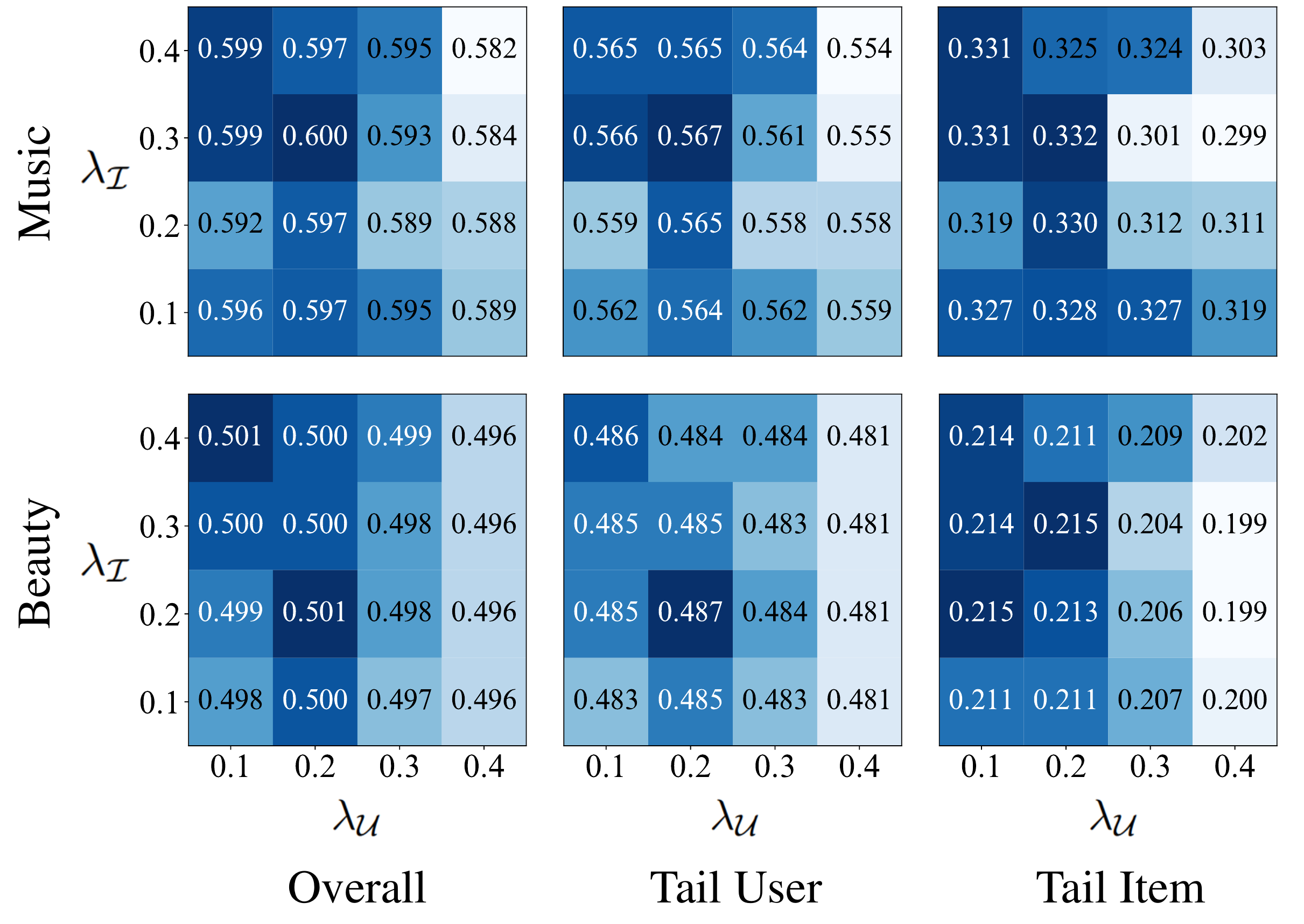} 
    \end{center}
    \vspace{-2.5ex}
    \caption{Sensitivity analysis of $\lambda_{\mathcal{U}}$ and $\lambda_{\mathcal{I}}$ (HR@10).}
    \label{fig:hyper}
    \vspace{-3.0ex}
\end{figure}

\subsection{Sensitivity Analysis}
\noindent\textbf{On $\lambda_{\mathcal{U}}$ and $\lambda_{\mathcal{I}}$ in Equation~\ref{eqn:eq6}. }
Figure~\ref{fig:hyper} shows the performance of overall, tail user, and tail item group according to the change of $\lambda_{\mathcal{U}}$ and $\lambda_{\mathcal{I}}$. 
We observe that the performance in the three groups shows similar trends over $\lambda_{\mathcal{U}}$ and $\lambda_{\mathcal{I}}$, e.g., $\lambda_{\mathcal{U}}=0.2$ and $\lambda_{\mathcal{I}}=0.3$  on Music data produces the best performance of overall, tail user, and tail item groups.
Thus, we can select the best model during training based only on the overall performance rather than tuning the hyper-parameters for all the specific groups, e.g., tail users and tail items,
which simplifies the tuning process of \proposed{}.
In addition, small values for $\lambda_{\mathcal{U}}$ and $\lambda_{\mathcal{I}}$ produce the best performance on all the groups compared with $L_{rec}$ in Equation \ref{eqn:eq6}, implying that the user and item branches act as regularizers of the recommendation loss, i.e., $\mathcal{L}_\text{rec}$.

\begin{figure}[t]
    \vspace{-1ex}
    \begin{center}
        \includegraphics[width=0.95\columnwidth]{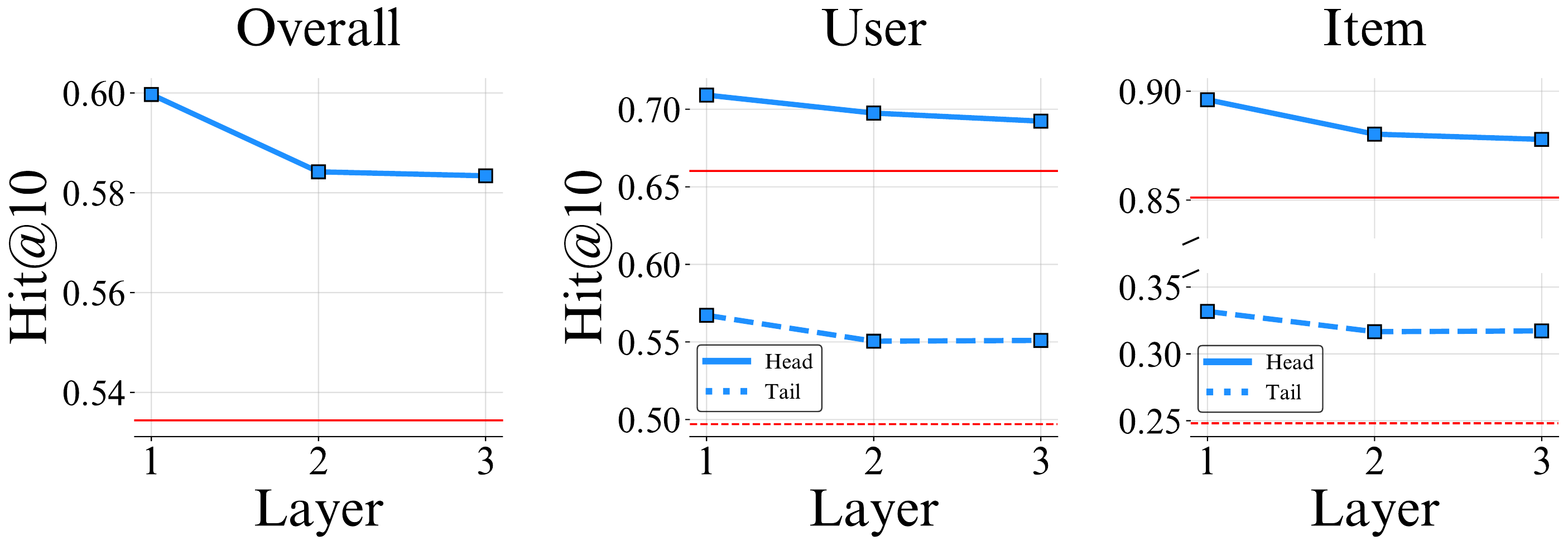} 
    \end{center}
    \vspace{-2.5ex}
    \caption{Performance of~\proposed~over different numbers of layers of $G_{\phi}^{\mathcal{U}}$ and $G_{\phi}^{\mathcal{I}}$ in Amazon Music data. Red lines indicate the performance of SASRec.}
    \label{fig:multi_layer}
    \vspace{-4.0ex}
\end{figure}

\smallskip
\noindent\textbf{On the model complexity of embedding generators. }
Figure~\ref{fig:multi_layer} shows the performance of~\proposed~over different numbers of layers of the embedding generators, i.e., $G_\phi^\mathcal{U}$ and $G_\phi^\mathcal{I}$. We observe that~\proposed~generally performs well even with a single-layer feed-forward neural network, which demonstrates that designing complex embedding generators is not beneficial.

\section{Related Works}

\noindent\textbf{Sequential Recommendation}. 
A myriad works in SRS have evolved based on modeling user's historical interaction with sequential dynamics. Recurrent neural networks (RNNs) have been successful in extracting the long-term interest of users \cite{gru4rec,hidasi2018recurrent,hidasi2016parallel}. 
Another line of the SRS model uses convolution neural networks (CNNs) to capture the sequential pattern by treating the item matrix as an image and applying the convolution operation \cite{caser,NextitNet}. 
Recently, the self-attention mechanism, which shows promising results in NLP domain \cite{transformer}, has been applied to SRS.
Specifically, SASRec \cite{sasrec} adopts the self-attention mechanism to capture the local and global interest, and following studies focus on improving  weaknesses of SASRec, e.g., heavy complexity of self-attention  \cite{lightweight,STOSA}, and dot-product for similarity computation \cite{STOSA}. BERT4Rec \cite{bert4rec} extends SASRec by incorporating both directions of a sequence along with prediction on masked items.  However, aforementioned works overlook the long-tailed problems, and thus show unsatisfactory performance on tail users and tail items.

\smallskip
\noindent\textbf{Long-Tailed Sequential Recommendation. }
Recent works in SRS have attempted to alleviate the long-tailed problem in terms of users or items. To address the long-tailed item problem, INSERT \cite{INSERT} leverages sequential information from similar users to supplement the lack of information of tail users. TP \cite{TP} utilizes the gradient alignment approach for minimizing the interference of the transfer process from head users to tail users, and adopts adversarial learning to map the head users and tail users in a shared latent space. Moreover, ASReP \cite{ASReP} augments previous item sequences to complement the short sequence of users, and reversely train the SRS model. DTSR \cite{DTSR} designs the item sequence as a probability distribution to explore the user's interest, and the interaction space to alleviate a short interest problem. 

To address the long-tailed item problem, CITIES \cite{CITIES} directly infers the tail item representation with item context using a self-attention aggregator, and freezes the parameters when inferring the tail item representation.
Tail-Net \cite{TailNet} distinguishes the sequence itself as a head or tail and adjusts the final prediction score. STOSA \cite{STOSA} replaces the dot-product of self-attention with Wasserstein distance to satisfy the triangle inequality among items, which implicitly alleviates the item cold-start problem. 

Although existing studies alleviate either the long-tailed user or long-tailed item problem, none of them jointly consider both problems. 
To the best of our knowledge, our work is the first to jointly consider both problems.

\looseness=-1
\section{Conclusion}

In this work, based on our empirical discovery that jointly addressing the long-tailed user and item problems is non-trivial with respect to tail users consuming the tail items (i.e., TT), we propose~\proposed{}, a novel framework for SRS that jointly alleviates the long-tailed user and item problems.~\proposed{} consists of bilateral branches each of which is responsible for alleviating the long-tailed user and item problems, where the branches mutually enhance each other along with the curriculum learning strategy in an end-to-end manner. 
The proposed framework is model-agnostic so that any existing SRS models can readily utilize it to alleviate the long-tailed user and item problems. 
Extensive experiments on \textit{eight} real-world datasets show the effectiveness of \proposed{} for tail users/items without scarifying the performance on head users/items.

\vspace{2ex}
\noindent\textbf{Acknowledgement}:
This work was supported by Institute of Information \& communications Technology Planning \& Evaluation (IITP) grant funded by the Korea government (MSIT) (No.2022-0-00157), and the National Research Foundation of Korea (NRF) grant funded by the Korea government(MSIT) (No.2021R1C1C1009081).

\bibliographystyle{ACM-Reference-Format}
\bibliography{MELT}


\begin{thebibliography}{32}


\ifx \showCODEN    \undefined \def \showCODEN     #1{\unskip}     \fi
\ifx \showDOI      \undefined \def \showDOI       #1{#1}\fi
\ifx \showISBNx    \undefined \def \showISBNx     #1{\unskip}     \fi
\ifx \showISBNxiii \undefined \def \showISBNxiii  #1{\unskip}     \fi
\ifx \showISSN     \undefined \def \showISSN      #1{\unskip}     \fi
\ifx \showLCCN     \undefined \def \showLCCN      #1{\unskip}     \fi
\ifx \shownote     \undefined \def \shownote      #1{#1}          \fi
\ifx \showarticletitle \undefined \def \showarticletitle #1{#1}   \fi
\ifx \showURL      \undefined \def \showURL       {\relax}        \fi
\providecommand\bibfield[2]{#2}
\providecommand\bibinfo[2]{#2}
\providecommand\natexlab[1]{#1}
\providecommand\showeprint[2][]{arXiv:#2}

\bibitem[Anderson(2006)]%
        {pareto}
\bibfield{author}{\bibinfo{person}{Chris Anderson}.}
  \bibinfo{year}{2006}\natexlab{}.
\newblock \bibinfo{booktitle}{\emph{The long tail: Why the future of business
  is selling less of more}}.
\newblock \bibinfo{publisher}{Hachette UK}.
\newblock


\bibitem[Bengio et~al\mbox{.}(2009)]%
        {bengio2009curriculum}
\bibfield{author}{\bibinfo{person}{Yoshua Bengio},
  \bibinfo{person}{J{\'e}r{\^o}me Louradour}, \bibinfo{person}{Ronan
  Collobert}, {and} \bibinfo{person}{Jason Weston}.}
  \bibinfo{year}{2009}\natexlab{}.
\newblock \showarticletitle{Curriculum learning}. In
  \bibinfo{booktitle}{\emph{Proceedings of the 26th annual international
  conference on machine learning}}. \bibinfo{pages}{41--48}.
\newblock


\bibitem[Box and Meyer(1986)]%
        {box1986analysis}
\bibfield{author}{\bibinfo{person}{George~EP Box} {and}
  \bibinfo{person}{R~Daniel Meyer}.} \bibinfo{year}{1986}\natexlab{}.
\newblock \showarticletitle{An analysis for unreplicated fractional
  factorials}.
\newblock \bibinfo{journal}{\emph{Technometrics}} \bibinfo{volume}{28},
  \bibinfo{number}{1} (\bibinfo{year}{1986}), \bibinfo{pages}{11--18}.
\newblock


\bibitem[Chen et~al\mbox{.}(2020)]%
        {chen2020esam}
\bibfield{author}{\bibinfo{person}{Zhihong Chen}, \bibinfo{person}{Rong Xiao},
  \bibinfo{person}{Chenliang Li}, \bibinfo{person}{Gangfeng Ye},
  \bibinfo{person}{Haochuan Sun}, {and} \bibinfo{person}{Hongbo Deng}.}
  \bibinfo{year}{2020}\natexlab{}.
\newblock \showarticletitle{Esam: Discriminative domain adaptation with
  non-displayed items to improve long-tail performance}. In
  \bibinfo{booktitle}{\emph{Proceedings of the 43rd International ACM SIGIR
  Conference on Research and Development in Information Retrieval}}.
  \bibinfo{pages}{579--588}.
\newblock


\bibitem[Fan et~al\mbox{.}(2021)]%
        {DTSR}
\bibfield{author}{\bibinfo{person}{Ziwei Fan}, \bibinfo{person}{Zhiwei Liu},
  \bibinfo{person}{Shen Wang}, \bibinfo{person}{Lei Zheng}, {and}
  \bibinfo{person}{Philip~S Yu}.} \bibinfo{year}{2021}\natexlab{}.
\newblock \showarticletitle{Modeling Sequences as Distributions with
  Uncertainty for Sequential Recommendation}. In
  \bibinfo{booktitle}{\emph{Proceedings of the 30th ACM International
  Conference on Information \& Knowledge Management}}.
  \bibinfo{pages}{3019--3023}.
\newblock


\bibitem[Fan et~al\mbox{.}(2022)]%
        {STOSA}
\bibfield{author}{\bibinfo{person}{Ziwei Fan}, \bibinfo{person}{Zhiwei Liu},
  \bibinfo{person}{Yu Wang}, \bibinfo{person}{Alice Wang},
  \bibinfo{person}{Zahra Nazari}, \bibinfo{person}{Lei Zheng},
  \bibinfo{person}{Hao Peng}, {and} \bibinfo{person}{Philip~S Yu}.}
  \bibinfo{year}{2022}\natexlab{}.
\newblock \showarticletitle{Sequential Recommendation via Stochastic
  Self-Attention}. In \bibinfo{booktitle}{\emph{Proceedings of the ACM Web
  Conference 2022}}. \bibinfo{pages}{2036--2047}.
\newblock


\bibitem[He et~al\mbox{.}(2016)]%
        {he2016vista}
\bibfield{author}{\bibinfo{person}{Ruining He}, \bibinfo{person}{Chen Fang},
  \bibinfo{person}{Zhaowen Wang}, {and} \bibinfo{person}{Julian McAuley}.}
  \bibinfo{year}{2016}\natexlab{}.
\newblock \showarticletitle{Vista: A visually, socially, and temporally-aware
  model for artistic recommendation}. In \bibinfo{booktitle}{\emph{Proceedings
  of the 10th ACM conference on recommender systems}}.
  \bibinfo{pages}{309--316}.
\newblock


\bibitem[Hidasi and Karatzoglou(2018)]%
        {hidasi2018recurrent}
\bibfield{author}{\bibinfo{person}{Bal{\'a}zs Hidasi} {and}
  \bibinfo{person}{Alexandros Karatzoglou}.} \bibinfo{year}{2018}\natexlab{}.
\newblock \showarticletitle{Recurrent neural networks with top-k gains for
  session-based recommendations}. In \bibinfo{booktitle}{\emph{Proceedings of
  the 27th ACM international conference on information and knowledge
  management}}. \bibinfo{pages}{843--852}.
\newblock


\bibitem[Hidasi et~al\mbox{.}(2015)]%
        {gru4rec}
\bibfield{author}{\bibinfo{person}{Bal{\'a}zs Hidasi},
  \bibinfo{person}{Alexandros Karatzoglou}, \bibinfo{person}{Linas Baltrunas},
  {and} \bibinfo{person}{Domonkos Tikk}.} \bibinfo{year}{2015}\natexlab{}.
\newblock \showarticletitle{Session-based recommendations with recurrent neural
  networks}.
\newblock \bibinfo{journal}{\emph{arXiv preprint arXiv:1511.06939}}
  (\bibinfo{year}{2015}).
\newblock


\bibitem[Hidasi et~al\mbox{.}(2016)]%
        {hidasi2016parallel}
\bibfield{author}{\bibinfo{person}{Bal{\'a}zs Hidasi}, \bibinfo{person}{Massimo
  Quadrana}, \bibinfo{person}{Alexandros Karatzoglou}, {and}
  \bibinfo{person}{Domonkos Tikk}.} \bibinfo{year}{2016}\natexlab{}.
\newblock \showarticletitle{Parallel recurrent neural network architectures for
  feature-rich session-based recommendations}. In
  \bibinfo{booktitle}{\emph{Proceedings of the 10th ACM conference on
  recommender systems}}. \bibinfo{pages}{241--248}.
\newblock


\bibitem[Hyun et~al\mbox{.}(2020)]%
        {hyun2020interest}
\bibfield{author}{\bibinfo{person}{Dongmin Hyun}, \bibinfo{person}{Junsu Cho},
  \bibinfo{person}{Chanyoung Park}, {and} \bibinfo{person}{Hwanjo Yu}.}
  \bibinfo{year}{2020}\natexlab{}.
\newblock \showarticletitle{Interest Sustainability-Aware Recommender System}.
  In \bibinfo{booktitle}{\emph{2020 IEEE International Conference on Data
  Mining (ICDM)}}. IEEE, \bibinfo{pages}{192--201}.
\newblock


\bibitem[Hyun et~al\mbox{.}(2022)]%
        {hyun2022beyond}
\bibfield{author}{\bibinfo{person}{Dongmin Hyun}, \bibinfo{person}{Chanyoung
  Park}, \bibinfo{person}{Junsu Cho}, {and} \bibinfo{person}{Hwanjo Yu}.}
  \bibinfo{year}{2022}\natexlab{}.
\newblock \showarticletitle{Beyond Learning from Next Item: Sequential
  Recommendation via Personalized Interest Sustainability}. In
  \bibinfo{booktitle}{\emph{Proceedings of the 31st ACM International
  Conference on Information \& Knowledge Management}}.
  \bibinfo{pages}{812--821}.
\newblock


\bibitem[Jang et~al\mbox{.}(2020)]%
        {CITIES}
\bibfield{author}{\bibinfo{person}{Seongwon Jang}, \bibinfo{person}{Hoyeop
  Lee}, \bibinfo{person}{Hyunsouk Cho}, {and} \bibinfo{person}{Sehee Chung}.}
  \bibinfo{year}{2020}\natexlab{}.
\newblock \showarticletitle{CITIES: Contextual Inference of Tail-Item
  Embeddings for Sequential Recommendation}. In \bibinfo{booktitle}{\emph{2020
  IEEE International Conference on Data Mining (ICDM)}}. IEEE,
  \bibinfo{pages}{202--211}.
\newblock


\bibitem[Jiang et~al\mbox{.}(2021)]%
        {arxiv}
\bibfield{author}{\bibinfo{person}{Juyong Jiang}, \bibinfo{person}{Yingtao
  Luo}, \bibinfo{person}{Jae~Boum Kim}, \bibinfo{person}{Kai Zhang}, {and}
  \bibinfo{person}{Sunghun Kim}.} \bibinfo{year}{2021}\natexlab{}.
\newblock \bibinfo{title}{Sequential Recommendation with Bidirectional
  Chronological Augmentation of Transformer}.
\newblock
\newblock
\urldef\tempurl%
\url{https://doi.org/10.48550/ARXIV.2112.06460}
\showDOI{\tempurl}
\showeprint[arxiv]{2112.06460}


\bibitem[Kang and McAuley(2018)]%
        {sasrec}
\bibfield{author}{\bibinfo{person}{Wang-Cheng Kang} {and}
  \bibinfo{person}{Julian McAuley}.} \bibinfo{year}{2018}\natexlab{}.
\newblock \showarticletitle{Self-attentive sequential recommendation}. In
  \bibinfo{booktitle}{\emph{2018 IEEE international conference on data mining
  (ICDM)}}. IEEE, \bibinfo{pages}{197--206}.
\newblock


\bibitem[Kim et~al\mbox{.}(2019)]%
        {kim2019sequential}
\bibfield{author}{\bibinfo{person}{Yejin Kim}, \bibinfo{person}{Kwangseob Kim},
  \bibinfo{person}{Chanyoung Park}, {and} \bibinfo{person}{Hwanjo Yu}.}
  \bibinfo{year}{2019}\natexlab{}.
\newblock \showarticletitle{Sequential and Diverse Recommendation with Long
  Tail}. In \bibinfo{booktitle}{\emph{IJCAI}}, Vol.~\bibinfo{volume}{19}.
  \bibinfo{pages}{2740--2746}.
\newblock


\bibitem[Li et~al\mbox{.}(2020)]%
        {li2020time}
\bibfield{author}{\bibinfo{person}{Jiacheng Li}, \bibinfo{person}{Yujie Wang},
  {and} \bibinfo{person}{Julian McAuley}.} \bibinfo{year}{2020}\natexlab{}.
\newblock \showarticletitle{Time interval aware self-attention for sequential
  recommendation}. In \bibinfo{booktitle}{\emph{Proceedings of the 13th
  international conference on web search and data mining}}.
  \bibinfo{pages}{322--330}.
\newblock


\bibitem[Li et~al\mbox{.}(2021)]%
        {lightweight}
\bibfield{author}{\bibinfo{person}{Yang Li}, \bibinfo{person}{Tong Chen},
  \bibinfo{person}{Peng-Fei Zhang}, {and} \bibinfo{person}{Hongzhi Yin}.}
  \bibinfo{year}{2021}\natexlab{}.
\newblock \showarticletitle{Lightweight self-attentive sequential
  recommendation}. In \bibinfo{booktitle}{\emph{Proceedings of the 30th ACM
  International Conference on Information \& Knowledge Management}}.
  \bibinfo{pages}{967--977}.
\newblock


\bibitem[Liu and Zheng(2020)]%
        {TailNet}
\bibfield{author}{\bibinfo{person}{Siyi Liu} {and} \bibinfo{person}{Yujia
  Zheng}.} \bibinfo{year}{2020}\natexlab{}.
\newblock \showarticletitle{Long-tail session-based recommendation}. In
  \bibinfo{booktitle}{\emph{Fourteenth ACM conference on recommender systems}}.
  \bibinfo{pages}{509--514}.
\newblock


\bibitem[Liu et~al\mbox{.}(2021)]%
        {ASReP}
\bibfield{author}{\bibinfo{person}{Zhiwei Liu}, \bibinfo{person}{Ziwei Fan},
  \bibinfo{person}{Yu Wang}, {and} \bibinfo{person}{Philip~S Yu}.}
  \bibinfo{year}{2021}\natexlab{}.
\newblock \showarticletitle{Augmenting sequential recommendation with
  pseudo-prior items via reversely pre-training transformer}. In
  \bibinfo{booktitle}{\emph{Proceedings of the 44th international ACM SIGIR
  conference on Research and development in information retrieval}}.
  \bibinfo{pages}{1608--1612}.
\newblock


\bibitem[Loshchilov and Hutter(2016)]%
        {loshchilov2016sgdr}
\bibfield{author}{\bibinfo{person}{Ilya Loshchilov} {and}
  \bibinfo{person}{Frank Hutter}.} \bibinfo{year}{2016}\natexlab{}.
\newblock \showarticletitle{Sgdr: Stochastic gradient descent with warm
  restarts}.
\newblock \bibinfo{journal}{\emph{arXiv preprint arXiv:1608.03983}}
  (\bibinfo{year}{2016}).
\newblock


\bibitem[McAuley et~al\mbox{.}(2015)]%
        {mcauley2015image}
\bibfield{author}{\bibinfo{person}{Julian McAuley},
  \bibinfo{person}{Christopher Targett}, \bibinfo{person}{Qinfeng Shi}, {and}
  \bibinfo{person}{Anton Van Den~Hengel}.} \bibinfo{year}{2015}\natexlab{}.
\newblock \showarticletitle{Image-based recommendations on styles and
  substitutes}. In \bibinfo{booktitle}{\emph{Proceedings of the 38th
  international ACM SIGIR conference on research and development in information
  retrieval}}. \bibinfo{pages}{43--52}.
\newblock


\bibitem[Song et~al\mbox{.}(2021)]%
        {INSERT}
\bibfield{author}{\bibinfo{person}{Wenzhuo Song}, \bibinfo{person}{Shoujin
  Wang}, \bibinfo{person}{Yan Wang}, {and} \bibinfo{person}{Shengsheng Wang}.}
  \bibinfo{year}{2021}\natexlab{}.
\newblock \showarticletitle{Next-item recommendations in short sessions}. In
  \bibinfo{booktitle}{\emph{Fifteenth ACM Conference on Recommender Systems}}.
  \bibinfo{pages}{282--291}.
\newblock


\bibitem[Sun et~al\mbox{.}(2019)]%
        {bert4rec}
\bibfield{author}{\bibinfo{person}{Fei Sun}, \bibinfo{person}{Jun Liu},
  \bibinfo{person}{Jian Wu}, \bibinfo{person}{Changhua Pei},
  \bibinfo{person}{Xiao Lin}, \bibinfo{person}{Wenwu Ou}, {and}
  \bibinfo{person}{Peng Jiang}.} \bibinfo{year}{2019}\natexlab{}.
\newblock \showarticletitle{BERT4Rec: Sequential recommendation with
  bidirectional encoder representations from transformer}. In
  \bibinfo{booktitle}{\emph{Proceedings of the 28th ACM international
  conference on information and knowledge management}}.
  \bibinfo{pages}{1441--1450}.
\newblock


\bibitem[Tang and Wang(2018)]%
        {caser}
\bibfield{author}{\bibinfo{person}{Jiaxi Tang} {and} \bibinfo{person}{Ke
  Wang}.} \bibinfo{year}{2018}\natexlab{}.
\newblock \showarticletitle{Personalized top-n sequential recommendation via
  convolutional sequence embedding}. In \bibinfo{booktitle}{\emph{Proceedings
  of the eleventh ACM international conference on web search and data mining}}.
  \bibinfo{pages}{565--573}.
\newblock


\bibitem[Vaswani et~al\mbox{.}(2017)]%
        {transformer}
\bibfield{author}{\bibinfo{person}{Ashish Vaswani}, \bibinfo{person}{Noam
  Shazeer}, \bibinfo{person}{Niki Parmar}, \bibinfo{person}{Jakob Uszkoreit},
  \bibinfo{person}{Llion Jones}, \bibinfo{person}{Aidan~N Gomez},
  \bibinfo{person}{{\L}ukasz Kaiser}, {and} \bibinfo{person}{Illia
  Polosukhin}.} \bibinfo{year}{2017}\natexlab{}.
\newblock \showarticletitle{Attention is all you need}.
\newblock \bibinfo{journal}{\emph{Advances in neural information processing
  systems}}  \bibinfo{volume}{30} (\bibinfo{year}{2017}).
\newblock


\bibitem[Yin et~al\mbox{.}(2012)]%
        {yin2012challenging}
\bibfield{author}{\bibinfo{person}{Hongzhi Yin}, \bibinfo{person}{Bin Cui},
  \bibinfo{person}{Jing Li}, \bibinfo{person}{Junjie Yao}, {and}
  \bibinfo{person}{Chen Chen}.} \bibinfo{year}{2012}\natexlab{}.
\newblock \showarticletitle{Challenging the long tail recommendation}.
\newblock \bibinfo{journal}{\emph{arXiv preprint arXiv:1205.6700}}
  (\bibinfo{year}{2012}).
\newblock


\bibitem[Yin et~al\mbox{.}(2020)]%
        {TP}
\bibfield{author}{\bibinfo{person}{Jianwen Yin}, \bibinfo{person}{Chenghao
  Liu}, \bibinfo{person}{Weiqing Wang}, \bibinfo{person}{Jianling Sun}, {and}
  \bibinfo{person}{Steven~CH Hoi}.} \bibinfo{year}{2020}\natexlab{}.
\newblock \showarticletitle{Learning transferrable parameters for long-tailed
  sequential user behavior modeling}. In \bibinfo{booktitle}{\emph{Proceedings
  of the 26th ACM SIGKDD International Conference on Knowledge Discovery \&
  Data Mining}}. \bibinfo{pages}{359--367}.
\newblock


\bibitem[Yuan et~al\mbox{.}(2019)]%
        {NextitNet}
\bibfield{author}{\bibinfo{person}{Fajie Yuan}, \bibinfo{person}{Alexandros
  Karatzoglou}, \bibinfo{person}{Ioannis Arapakis}, \bibinfo{person}{Joemon~M
  Jose}, {and} \bibinfo{person}{Xiangnan He}.} \bibinfo{year}{2019}\natexlab{}.
\newblock \showarticletitle{A simple convolutional generative network for next
  item recommendation}. In \bibinfo{booktitle}{\emph{Proceedings of the twelfth
  ACM international conference on web search and data mining}}.
  \bibinfo{pages}{582--590}.
\newblock


\bibitem[Yuan et~al\mbox{.}(2014)]%
        {yuan2014graph}
\bibfield{author}{\bibinfo{person}{Quan Yuan}, \bibinfo{person}{Gao Cong},
  {and} \bibinfo{person}{Aixin Sun}.} \bibinfo{year}{2014}\natexlab{}.
\newblock \showarticletitle{Graph-based point-of-interest recommendation with
  geographical and temporal influences}. In
  \bibinfo{booktitle}{\emph{Proceedings of the 23rd ACM international
  conference on conference on information and knowledge management}}.
  \bibinfo{pages}{659--668}.
\newblock


\bibitem[Yun et~al\mbox{.}(2022)]%
        {yun2022lte4g}
\bibfield{author}{\bibinfo{person}{Sukwon Yun}, \bibinfo{person}{Kibum Kim},
  \bibinfo{person}{Kanghoon Yoon}, {and} \bibinfo{person}{Chanyoung Park}.}
  \bibinfo{year}{2022}\natexlab{}.
\newblock \showarticletitle{LTE4G: Long-Tail Experts for Graph Neural
  Networks}. In \bibinfo{booktitle}{\emph{Proceedings of the 31st ACM
  International Conference on Information \& Knowledge Management}}.
  \bibinfo{pages}{2434--2443}.
\newblock


\bibitem[Zhou et~al\mbox{.}(2022)]%
        {FMLP}
\bibfield{author}{\bibinfo{person}{Kun Zhou}, \bibinfo{person}{Hui Yu},
  \bibinfo{person}{Wayne~Xin Zhao}, {and} \bibinfo{person}{Ji-Rong Wen}.}
  \bibinfo{year}{2022}\natexlab{}.
\newblock \showarticletitle{Filter-enhanced MLP is all you need for sequential
  recommendation}. In \bibinfo{booktitle}{\emph{Proceedings of the ACM Web
  Conference 2022}}. \bibinfo{pages}{2388--2399}.
\newblock


\end{thebibliography}

\end{document}